\documentclass[english,a4paper, twoside]{article}
\usepackage[T1]{fontenc}
\usepackage[latin9]{inputenc}
\usepackage{float}
\usepackage{bm}
\usepackage{amsmath}
\usepackage{graphicx}

\makeatletter

\usepackage{float}

\usepackage{bm}

\makeatletter

\usepackage{geometry}

\usepackage{esint}

\renewcommand{\i}{\mathrm{i}}
\renewcommand{\d}{\mathrm{d}}
\newcommand{\e}{\mathrm{e}}

\newcommand{\reffig}[1]{Figure~\ref{#1}}



\makeatother

\makeatother

\usepackage{babel}

\begin{document}

\title{Large time behavior and asymptotic stability of the two-dimensional
Euler and linearized Euler equations}

\author{Freddy Bouchet$^{*,**}$ and Hidetoshi Morita$^{*}$}

\maketitle
Affiliations : $*$ INLN, CNRS, UNSA, 1361 route des lucioles, 06
560 Valbonne, France \\
$**$ CNLS, Los Alamos National Laboratory, PO Box 1663, Los Alamos, NM 87545, United States

Corresponding author : Freddy Bouchet ; email : Freddy.Bouchet@inln.cnrs.fr
\\


\begin{abstract}
We study the asymptotic behavior and the asymptotic stability of the
two-dimensional Euler equations and of the two-dimensional linearized
Euler equations close to parallel flows. We focus on flows with spectrally stable
profiles $U\left(y\right)$ and with stationary streamlines $y=y_{0}$
(such that $U'(y_{0})=0$), a case that has not been studied previously.
We describe a new dynamical phenomenon: the depletion of the vorticity
at the stationary streamlines. An unexpected consequence, is that
the velocity decays for large times with power laws, similarly to
what happens in the case of the Orr mechanism for base flows without
stationary streamlines. The asymptotic behaviors of velocity and the
asymptotic profiles of vorticity are theoretically predicted and compared
with direct numerical simulations. We argue on the asymptotic stability
of this ensemble of flow profiles even in the absence of any dissipative
mechanisms.
\end{abstract}

\section{Introduction}

The flow of a perfect fluid is described by the Euler equations, one
of the oldest equations of mathematical physics \cite{euler250}.
Four century after their discovery by Euler, these equations still
propose big challenges both to mathematician and physicists \cite{euler250}.
Two-dimensional flows and the two-dimensional Euler equations are
mathematically much simpler than their three dimensional counterparts,
but still present some very interesting unsolved problems. One of
the main phenomena for two-dimensional flows is the self-organization
into coherent structures \cite{Miller:1990_PRL_Meca_Stat,Robert:1991_JSP_Meca_Stat,Marteau_Cardoso_Tabeling_1995PhRvE,Sommeria_1986_JFM_2Dinverscascade_MHD,Schneider_Farge_2008PhysicaD,Maassen_Clercx_VanHeijst_2003JFM}: monopoles, dipoles, and parallel flows. Such large-scale structures
are analogous to geophysical cyclones, anticyclones, and jets in the
ocean and atmospheres. This analogy, understood thanks to the theoretical
strong similarities between the 2D Euler equations on one hand and
the Quasi-Geostrophic or the Shallow Water models on the other hand,
is one of the main motivations for the study of the 2D Euler equations.
The 2D Euler equations also describe experimental flows: the transverse
dynamics of electron plasma columns \cite{Schecter_Dubin_etc_Vortex_Crystals_2DEuler1999PhFl},
the dynamics of fluids when three-dimensional motion is prevented
by a strong transverse field (rotation, a transverse magnetic field
in a liquid metal, etc. \cite{Sommeria_1986_JFM_2Dinverscascade_MHD})
or the dynamics of fluids in very thin geometries \cite{Paret_Tabeling_1998_PhysFluids}.

Because large-scale coherent flows appear spontaneously in two dimensional turbulence\textbf{, 
}their stability is a crucial problem. Moreover, the study of the
dynamical mechanism that describes 
the relaxation towards these stable flows is essential. 
In this paper, we consider stable parallel base flows $\mathbf{v}_{0}(x,y)=U(y)\mathbf{e}_{x}$, which are dynamical equilibria of the 2D-Euler equations.
We prove that the velocity of these flows is \emph{asymptotically stable}, that is that all the solutions to the nonlinear Euler equations
that start near $\mathbf{v}_{0}$ converge to some other parallel
flows $\mathbf{v}_{0}+\delta U\left(y\right)\mathbf{e}_{x}$, close
to $\mathbf{v}_{0}$\footnote{This notion of convergence makes the notion of asymptotic stability stronger
than the alternative Lyapounov stability, that only states that all solutions
of the nonlinear equations that start near a steady point $\mathbf{v_{0}}$
stay near $\mathbf{v_{0}}$ forever. The asymptotic stability of the
velocity refers to asymptotic stability in the kinetic energy norm. %
}. Our analysis mostly relies on the linearization
of the Euler equations close to the base flows, which we prove to actually
describe also the nonlinear relaxation at leading order. More precisely, we prove that
the perturbation velocity decays algebraically for large times. As far as the linearized dynamics is concerned, an important improvement
over the previous works is the understanding of the case when the
flow has some stationary streamlines $y=y_0$ (or equivalently the velocity profile $U\left(y\right)$ has some stationary points $y_{0}$,
$U'\left(y_{0}\right)=0$)\footnote{Please note that the profile $U(y)$ has stationary points, but the two-dimensional base flow has no stationary points}%
, which has not been elucidated even qualitatively
previously.

Besides the stability and the asymptotic stability problem itself,
the evolution operator for the linearized Euler equations plays a very
important role in different statistical approaches to turbulent flows
\cite{Dubin_ONeil_1988_PhysRevLett_Kinetic_Point_Vortex,Chavanis_houches_2002,Farrell_Ioannou_2003JAtS}.
Indeed, in a turbulent context, it is likely that the qualitative
or quantitative properties of the fluctuations around such stable
structures are related to the linearized dynamics. For instance, in
quasi-linear approaches or second order closure of the evolution of
the Euler equations, point vortex model \cite{Dubin_ONeil_1988_PhysRevLett_Kinetic_Point_Vortex},
or of the Navier-Stokes equation (either forced or unforced, and either
deterministic or stochastic,) the linear operator appears naturally
as an essential theoretical tool. Similarly, in the forced problems,
in the linear regime, the response can be easily expressed in terms
of the evolution operator for the linearized dynamics or of the resolvent
operator for the dynamics. The behavior of such operators for large
times is thus a very important issue that has many theoretical and
dynamical consequences. In this work, we quantify very precisely the
large time asymptotic behavior of the evolution operator for the linearized
2D-Euler equations, and discuss briefly some of the implications for
the above problems.\\

The stability of the large-scale coherent structures of two-dimensional
flows is a very old and classical field of fluid mechanics. For instance
Rayleigh \cite{Rayleigh_1878}, Kelvin \cite{Kelvin_1887}, Orr \cite{Orr_1907},
Sommerfeld \cite{Sommerfeld_1908} and many other famous scientists
from the nineteenth and the beginning of the twentieth century have
participated to the understanding of the linear theory for the 2D
Euler equations close to parallel flows. Mathematicians gave also
important contributions: Arnold's theorems \cite{Arnold_1966} and
some modern generalizations \cite{Wolansky_Ghil_1998_CMaPh,Simonnet_2008PhyD,EllisHavenTurkington:2002_Nonlinearity_Stability,Caglioti_Pulvirenti_Rousset_200_8JPhA,Bouchet:2008_Physica_D}
prove the Lyapounov stability of some of these flows. Even if this work deals
only with the behavior of slightly perturbed stable flows, other equally
interesting and important problems arise in the study of unstable
or oscillatory flows. Recently many works have been devoted to the
proof of the instability of some classes of flows, the characterization
of the spectrum of the linearized equations, and some estimates on
the stability and the instability of such flows ; see for instance
\cite{Friedlander_Howard_2005_Stud_Appl_Math,Shvydkoy_Friedlander_2002_Contemp_math,Grenier_Jones_Rousset_2005_Nonlinearity,Belenkaya_Friedlander_Yudovich_1999}
and references therein.

Our work is based on the linearized 2D Euler equations. The Rayleigh
equation \cite{Rayleigh_1878}, which describes modes for the linear
dynamics, has been a subject of mathematical and theoretical research
since the beginning of the twentieth century \cite{Drazin_Reid_1981},
and is still currently active. The main interest lies in the dynamical
phenomena associated with the singularities at the critical layers
(the singularities appearing when the frequency of the perturbation
is equal to that of a closed streamline of the base flow). However,
the modes of the Rayleigh equation do not describe the fully linearized
dynamics, because the linear operator is non-normal \cite{Farrell_Ioannou_1996_JAtS}.
Among the peculiarities of the linearized Euler equations, we stress
the Orr mechanism \cite{Orr_1907}: the base flow shears the perturbation
producing thinner and thinner filaments ; then when the velocity or
the streamfunction is computed, the effect of such filaments being
smoothed out, the perturbation velocity decays for large times. This
mechanism is easily quantified when the shear is linear, for the Euler equations \cite{Orr_1907}
or using for instance Kelvin waves \cite{Kelvin_1887} for viscous
flows. Case \cite{Case_1960_Phys_Fluids} and Dikii \cite{Dikii_1960}
were the first to stress that in general, for inviscid flows, such
a phenomenon is outside the scope of a modal description using the
Rayleigh equation. When the shear is linear, using a Fourier-Laplace
transform, 
the dynamics of the perturbation is properly described in the framework
of an initial value problems. They concluded that the perturbation
velocity decreases asymptotically with an algebraic law for large
times. Other phenomena associated to the non-normality of the linear
operator include possible transient growth \cite{Farrell_1987_JAS,Volponi_2005_JPhysA,Antkowiak_Brancher_2004_PhsFluids},
inviscid damping (the counterpart of Landau damping in plasmas), axisymmetrization
\cite{Scecter_etal_2000_PhysicsFluids,LeDizes_JFM_2000JFM}, and algebraic
instabilities \cite{Nolan_Montgomery_2000_JAtS}. From a mathematical
point of view, the singularities at the critical layers lead to the
existence of a continuous spectrum for the linearized Euler equation.
The analysis of the properties of this continuous spectrum explains
most of these transient growth, inviscid damping, algebraic instabilities,
and so on.

From a theoretical point of view, one class of works used the Laplace
transform tools \cite{Balmforth_Morrison_1999,Balmforth_Del-Castillo-Negrete_Young_1997_JFM,Schecter_Dubin_etc_Vortex_Crystals_2DEuler1999PhFl},
following the initial works of Case \cite{Case_1960_Phys_Fluids},
Dikii \cite{Dikii_1960} and the generalization to non-uniform shear
by Briggs-Daugherty and Levy \cite{Briggs_BDL_1970_Phys_Fluids}.
Another class of studies, less general but very enlightening, used
simple or particular base flows or special conditions for which explicit
computations are possible \cite{Kelvin_1887,Orr_1907,Yamagata_1976_JPO,Farrell_1987_JAS,Tung_1983_JFM,Brunet_Haynes_1995_JAtS,Brunet_Montgomery_2002_DynAtmOce,Brunet_Warn_1990_JAtS,Nolan_Montgomery_2000_JAtS}.

In this paper, we are especially interested in the precise description
of the large time asymptotic behavior of the 2D Euler and 2D linearized
Euler equation close to parallel base flows. For the linearized dynamics of stable base flows,
once the contribution of possible neutral modes has been subtracted,
the asymptotic behavior is related to the continuous spectrum of the
linearized operator. In the case of the base flow with a linear shear,
$U\left(y\right)=\sigma y$, the explicit computations by Case \cite{Case_1960_Phys_Fluids}
showed that, due to the Orr mechanism, for large times the velocity
perturbation decays algebraically : \begin{equation}
v_{x}\underset{t\rightarrow\infty}{\sim}\frac{C\left(y\right)}{t^{\alpha}}\,\,\,\text{and}\,\,\, v_{y}\underset{t\rightarrow\infty}{\sim}\frac{C\left(y\right)}{t^{\beta}},\label{eq:Vitesse_Asymptotics}\end{equation}
 with exponents $\alpha=1$ for the longitudinal component $v_{x}$
of the velocity perturbation and $\beta=2$ for the transverse one
$v_{y}$.

For more general base flows with strictly monotonic profiles $U\left(y\right)$ (without stationary streamline), it is a common belief
that the exponents $\alpha=1$ and $\beta=2$ remain valid. This belief is based on the results of
an ansatz for large time asymptotics \cite{Brown_Stewartson_1980_JFM} (see also \cite{Lundgren_1982_PhFl}, appendix A).  Some interesting comments about the temporal behavior of the streamfunction and velocity, in the case of localized initial perturbations (vorticity defects), can also be found in  \cite{Balmforth_Del-Castillo-Negrete_Young_1997_JFM} section 7. Even if we have not found any complete rigorous proofs, very precise classical arguments
using the Laplace transform \cite{Rosencrans_Sattinger_1966_J_Math_Phys,Briggs_BDL_1970_Phys_Fluids} conclude that the contribution of
the continuous spectrum to the stream function perturbation $\psi$ decays with $\psi\underset{t\rightarrow\infty}{=}\mathcal{O}\left(\frac{1}{t}\right)$, in agreement with equation (\ref{eq:Vitesse_Asymptotics}). However, these arguments do not generalize where the profile $U(y)$ is not monotonic (flows with stationary streamlines).

From Lundgren work (\cite{Lundgren_1982_PhFl}, appendix A), we see
that the preceding algebraic decay for the velocity or the streamfunction may be related
to the following asymptotic behavior for the perturbation vorticity: \begin{equation}
\omega\left(y,t\right)\underset{t\rightarrow\infty}{\sim}\omega_{\infty}\left(y\right)\exp\left(-ikU(y)t\right)+\mathcal{O}\left(\frac{1}{t^{\gamma}}\right),\label{eq:Lungren}\end{equation}
 where $k$ is the initial perturbation wave number. Indeed, computing
the velocity from Lundgren's ansatz (equation (\ref{eq:Lungren}))
and assuming uniformity in the asymptotic expansion, we obtain oscillating
integrals leading to algebraic decay for large times. The values for
the exponents $\alpha=1$ and $\beta=2$ are then related to the singularities
of the Green function used in order to compute the velocity perturbation
from the vorticity perturbation. This argument, assuming Lundgren's
ansatz, suggests that the asymptotic behavior for the velocity should
be different for velocity profiles $U(y)$ with stationary points
$y_{0}$ ($U'\left(y_{0}\right)=0$, base flow with stationary streamline). Actually, in such a case, the
stationary phase asymptotics for oscillating integrals would generically
give $1/\sqrt{t}$ contributions. It has then been noticed by several authors,
that with such a $1/\sqrt{t}$ law, Lundgren's ansatz would be self-consistent. \cite{Brown_Stewartson_1980_JFM,Brunet_Warn_1990_JAtS,Brunet_Haynes_1995_JAtS}. Similar problems have also been noticed by Brown and Stewartson \cite{Brown_Stewartson_1980_JFM},
as their own asymptotic expansion clearly breaks down where $U'(y)=0$ (base flows with stationary streamlines). Besides Lundgren
and Stewartson, many authors have insisted on the specificity of base flows with stationary streamlines (see for instance \cite{Brunet_Warn_1990_JAtS,Brunet_Haynes_1995_JAtS,Isichenko_1997_PhRvL}).\\

In the past, there have been only a few studies considering base flows with stationary streamlines. In the case of the equations for
2D barotropic flows on a $\beta$ plane (a direct generalization of
the 2D Euler equations), Brunet and coauthors \cite{Brunet_Warn_1990_JAtS,Brunet_Haynes_1995_JAtS}
have studied the dynamics close to a parabolic jet when the potential vorticity
gradient exactly cancels the $\beta$ effect. This case 
is similar to the linear shear case in the Euler equations studied
by Kelvin, Orr, Case and others \cite{Kelvin_1887,Orr_1907,Case_1960_Phys_Fluids,Tung_1983_JFM,Farrell_1987_JAS},
in that the vorticity gradient exactly cancels out, which makes the
linearized equation much simpler and amenable to a very interesting explicit analytic
treatment. In the following, we will argue that, because of the cancelation
of the vorticity (or potential vorticity) gradient, the dynamics of these cases is actually
non generic, and that flows where the vorticity (or potential vorticity) gradient does not vanish behave differently.

In the general case, the asymptotic behavior of the vorticity and velocity perturbations
of flows with stationary streamlines thus remains unstudied.  In natural flows, however, jet velocity profiles are most of time
not monotonic but have some extrema, i.e. the flow has stationary streamlines %
; see for instance Jupiter, atmospheric, and ocean jets.
Why have cases with stationary streamlines not been studied previously ? This may be
partially on account of the wrong belief that base velocity profiles
with stationary points should be unstable. It is true that many of the flows with extrema in their velocity profiles
do not fulfill the classical Rayleigh-Fjørtoft criteria \cite{Drazin_Reid_1981}.
However, these criteria are only sufficient conditions of stability.
Moreover, as seen in natural flows and as shown bellow with several examples, many
parallel flows with stationary streamlines and not fulfilling the classical Rayleigh-Fjørtoft criteria are
actually stable.
Another reason for a lack of studies may also be the theoretical difficulty with Laplace tools in this case,
related to the presence of stationary streamlines (merging of critical
layers). \textbf{
}Indeed, an essential tool for the Laplace transform is the analytic
continuation of dynamical quantities, performed by avoiding the singularities
associated to the critical layers, with the use of integration in
the complex plane \cite{Drazin_Reid_1981}.
As will be discussed bellow, in the case of flows with stationary
streamlines, in order to perform the analytic continuation, one would need to find a path in the complex plane  passing at the same time above and bellow the singularity, which is clearly impossible. For this reason,
it has often been stated that Laplace tools cannot be used when stationary streamlines are present. By contrast, we illustrate in this work that even if analytic continuation can not been performed in this way, Laplace tools are
still very useful and lead to very interesting results.

\begin{figure}
\begin{center}
\includegraphics[width=0.9\textwidth]{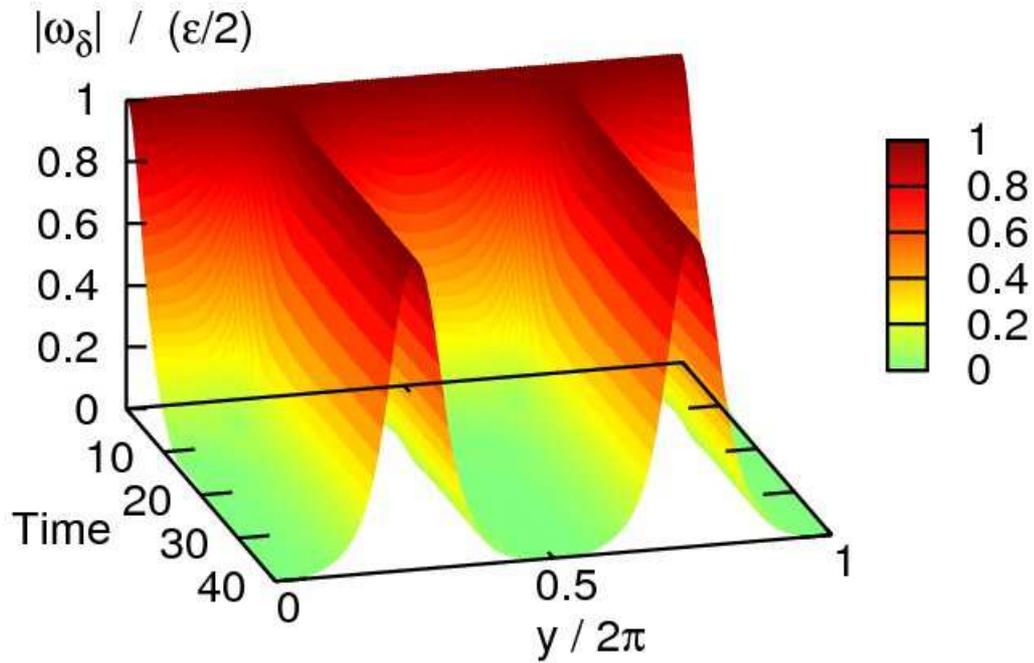}
\end{center}
\caption{Evolution of the vorticity perturbation
$\omega(x,y,t)=\omega\left(y,t\right)\exp\left(ikx\right)$, close
to a parallel flow $\mathbf{v}_{0}(x,y)=U(y)\mathbf{e}_{x}$ with
$U\left(y\right)=\cos\left(y\right)$. The figure shows the modulus
of the perturbation $\left|\omega\left(y,t\right)\right|$ as a function
of time and $y$. One clearly sees that the vorticity perturbation
rapidly converges to zero close to the points where the velocity profile
$U\left(y\right)$ has extrema ($y=0$ and $\pi$). This \emph{depletion of the perturbation
vorticity at the stationary streamlines} is a new generic self-consistent mechanism, understood
mathematically as the regularization of the critical layer singularities
at the edge of the continuous spectrum.}
\label{fig:Vorticity_Depletion}
\end{figure}

In the following we consider the generic case of a parallel
flow with any profile, either with or without stationary streamlines,
improving by far the class of previously studied flows and overcoming the previously discussed difficulties. We also discuss
possible generalizations to monopole vortices. We show how the Laplace transform is generalized to
the case of base flows with stationary streamlines. For instance, we show
how the classical determination of the number of unstable modes, by
using Nyquist's plots, remains valid in this case. From this general
theoretical approach, we prove that the asymptotic vorticity field
actually follows the Lundgren's ansatz (\ref{eq:Lungren}), even in
the case of a base flow with stationary streamlines. Similarly the velocity
field decreases also algebraically with the power laws (\ref{eq:Vitesse_Asymptotics}),
with $\alpha=1$ and $\beta=2$%
\footnote{An exception is the velocity field close to the stationary streamline,
where we have no theoretical prediction, but where we observe numerically
that either $\alpha=1$ and $\beta=2$ or $\alpha=\beta=3/2$ depending
on the symmetry of the perturbation.%
}. This may seem paradoxical, after the discussion of the preceding
paragraphs. Actually, the naturally expected $1/\sqrt{t}$ contributions
from the stationary phase asymptotics do not exist, unexpectedly. One reason is the
non-uniformity of the asymptotic expansion in Lundgren's ansatz. Another
more important reason is related to a new dynamical phenomena leading
to the rapid decrease and cancelation of the vorticity perturbation
exactly at the stationary streamline (see Fig. \ref{fig:Vorticity_Depletion}),
which partially erase the effect of the stationary phase. We call
this phenomena \emph{vorticity depletion at the stationary streamlines}.
This is a non-local collective phenomena, due to the effect of the
perturbation velocity on the background vorticity gradient. For this
reason, this phenomena has not been observed in the previous studies
involving stationary streamlines \cite{Brunet_Warn_1990_JAtS,Brunet_Haynes_1995_JAtS,Isichenko_1997_PhRvL},
because these cases have an exactly zero vorticity (or potential vorticity) gradient. These last cases are thus non generic. \\

In the following, we predict the vorticity depletion at the stationary
streamlines using Laplace tools. It is thus a generic effect in any
type of parallel flow with non-monotonic stable velocity profile. We also illustrate the results
by direct numerical simulations in the case of Kolmogorov base flow
$U(y)=\cos{y}$, for the 2D Euler equations with periodic boundary
conditions.
This vorticity depletion mechanism also impacts non linear turbulent flows when
the perturbations are small enough to be governed by the linearized equations,
as discussed in the conclusion.

We establish the large time asymptotic contributions to the vorticity and to the
velocity fields,  by the continuous spectrum, in the case of  the linearized dynamics. We are then able to discuss
the asymptotic stability of the velocity of parallel base flows, for the non-linear dynamics. The
result is that parallel base flows which have no mode (neither stable
nor unstable) are asymptotically stable: any small perturbation leads
to a small deformation of the base flow, the perturbation velocity
to this new base flow decaying algebraically. Stable Kolmogorov flows
are examples of base flows without modes, illustrating the importance
of this class of flows. Based on these results we also conclude that a quasi-linear approach predicts the asymptotic velocity profile. We note that all this is true only thanks to a non-trivial cancelation of leading order terms, already noted by \cite{Lundgren_1982_PhFl}. The current work put the validity of a quasi-linear approach on a more rigorous ground, and proves that this it is also valid also for flows with stationary streamlines. It also gives an efficient theory and numerical tool to predict the asymptotic flow. \\

For the case of a circular vortex base flows, the work \cite{Bassom_Gilbert_JFM_1998} shows that the far field velocity decays with exponents different from the case of parallel flows. For circular vortex base flows, an example of non-monotonic angular velocity profile has also been studied \cite{Nolan_Montgomery_2000_JAtS}, based on a special explicit solution \cite{Smith_Rosenbluth_1990_PRL}.
This example shows a very interesting algebraic instability with $t^{1/2}$
growths. Even if we do not explicitly treat the case of stable
circular vortices $\mathbf{v}\left(r\right)=U\left(r\right)\mathbf{e}_{\theta}$
in the present study, the generalization to this case of the present study could be taken
following similar theoretical arguments. The mechanism of vorticity depletion at stationary streamlines or at the core of stable probably also exist for stable vortices, as discussed in the conclusion. \\

In section \ref{sec:theory} we introduce the 2D Euler equations and
the linearized Euler equations. Section \ref{sec:theory} describes
the theory related to the linearized Euler equation. We discuss the
main results related to the asymptotic behavior of the vorticity and velocity
fields in section \ref{sub:Large-time-asymptotic}. The core of the
proof relies on the results of the analysis on the limit of small
$\epsilon$ for the resolvent operator in section \ref{sub:resolvent}.

Section \ref{sec:Asymptotic-stability} discusses the asymptotic stability
of parallel flows for the Euler equation.

These results are illustrated in the case of doubly periodic boundary
conditions, with the Kolmogorov flows as a base flow, in section \ref{sec:Kolmogorov-flow}.
For some aspect ratio, these flow are stable even if they do not fulfill
the hypothesis of any of the two Arnold's theorems. Applying Arnold's
ideas to this case, we first prove their Lyapounov stability in section
\ref{sec:Stability}. In section \ref{sub:numerical-computation},
we show the results of direct numerical simulations of the 2D Euler
equations (nonlinear), which both illustrate the theoretical results
of section \ref{sec:theory}, and show that the linearized dynamics
correctly describes the nonlinear one.

Section \ref{sec:Discussion}
discusses briefly some consequences of these results for the 2D Euler
and Navier-Stokes equations with stochastic forces and for possible
theories relying on a quasi-linear or kinetic approach. It also discusses
some possible generalizations to models of interest for geophysical
flows.

\section{The 2D Euler and linearized Euler equations\label{sec:theory}}

Let us consider the 2D Euler equations \begin{equation}
\frac{\partial\Omega}{\partial t}+\mathbf{V}.\mathbf{\nabla}\Omega=0,\label{eq:Euler}\end{equation}
 where $\Omega$ is the vorticity and $\mathbf{V}$ is the velocity.
We consider this equation either in an infinite plane, in a channel
geometry with boundary conditions $\mathbf{V.n}=0$ on the boundary
wall, or on a doubly periodic domain $(0,2\pi/\delta)\left(0,2\pi\right)$,
where $\delta>1$ is the aspect ratio. In some parts of the discussion,
for technical reasons the boundary conditions will be important. Then
only the case of a doubly periodic domain will be explicitly treated.
However all the results are applicable to the channel and infinite
domain geometries with slight modifications.

We study the asymptotic stability of parallel flows $\mathbf{v}_{0}$=$U\left(y\right)\mathbf{e}_{x}$.
We will thus consider the Euler equations (\ref{eq:Euler}) with initial
conditions close to this base flow : $\Omega=\omega_{0}+\omega$ and
$\mathbf{V}=\mathbf{v}+\mathbf{v}_{0}$, where $\omega_{0}\left(y\right)=-U'\left(y\right)$
is the base flow vorticity and $\omega$ and $\mathbf{v}$ are the
perturbation vorticity and velocity, respectively.

We also need to consider the linearized 2D Euler equations close to this
base flow. It reads

\begin{equation}
\frac{\partial\omega}{\partial t}+\mathbf{v}.\mathbf{\nabla}\omega_{0}+\mathbf{v}_{0}.\mathbf{\nabla}\omega=0.\label{eq:Linearized_Euler}\end{equation}
 \\

We assume that the base flow $U\left(y\right)$ has no unstable mode
(a precise definition of modes will be given along the discussion).
In section \ref{sec:Kolmogorov-flow} we will illustrate some of the
results on the particular case of the Kolmogorov flow $U(y)=\cos{y}$
(in a doubly periodic domain).

\subsection{The Laplace transform, resolvent operator and Rayleigh equation\label{sub:Initial_value_problems}}

In this section and the following we consider the linearized 2D Euler
equations. We give the main definitions used later on.

We decompose the perturbation vorticity in Fourier series for the
$x$ variable only. For parallel flows, due to the translational invariance,
these Fourier modes are independent one from the others for the linear
dynamics. In the following, we thus study perturbations of the form
$\omega(x,y,t)=\omega_{k}\left(y,t\right)\exp\left(ikx\right)$ and
$\psi(x,y,t)=\psi_{k}\left(y,t\right)\exp\left(ikx\right)$, where
$\psi$ is the stream function, with $\omega=\Delta\psi$ and $k$
is the longitudinal wave number. In the following, we drop the $k$
subscripts for the perturbation. The relations between $\omega$,
$\mathbf{v}$ and $\psi$ are then \begin{equation}
\omega=\frac{d^{2}\psi}{dy^{2}}-k^{2}\psi,\,\,\, v_{x}=-\frac{d\psi}{dy}\,\,\,\mbox{and}\,\,\, v_{y}=ik\psi.\label{eq:relation_omega_v}\end{equation}

The linearized Euler equations then reads \begin{equation}
\frac{\partial\omega}{\partial t}+ikU\left(y\right)\omega-ik\psi U''\left(y\right)=0\label{eq:Linearized_Euler_Fourier}\end{equation}

We will study the long time asymptotics of the linearized equation.
The more general approach is to use the Laplace transform of equation
(\ref{eq:Linearized_Euler_Fourier}). We define the Laplace transform
$\hat{\omega}$ of $\omega$ as \[
\hat{\omega}\left(y,p\right)=\int_{0}^{\infty}dt\,\omega\left(y,t\right)\exp\left(-pt\right)\]
 The Laplace transform is analytic for any complex number $p$ for
sufficiently large real part $\Re p$. The inverse Laplace transform
is given by \begin{equation}
\omega\left(y,t\right)=\frac{1}{2\pi i}\int_{\Gamma}dp\,\hat{\omega}\left(y,p\right)\exp\left(pt\right)\label{eq:inverse_Laplace}\end{equation}
 where the complex integration is performed along a Bromwich contour
$\Gamma$ in the complex plane of $p$. In the following we use the
notation $p=-ik\left(c+i\epsilon\right)$ where $c$ and $\epsilon$
are real numbers ; $c$ and $\epsilon$ are homogeneous to velocities.
We assume $k>0.$ The Laplace transform $\hat{\omega}$ is thus analytic
for sufficiently large $\epsilon$.

The Laplace transform of equation (\ref{eq:Linearized_Euler_Fourier})
reads \begin{equation}
\left(U(y)-c-i\epsilon\right)\hat{\omega}-U''\left(y\right)\phi=\frac{\omega\left(y,0\right)}{ik},\label{eq:Resolvent_vorticity}\end{equation}
 where $\phi=\hat{\psi}$ is the Laplace transform of $\psi$ and
$\omega\left(y,0\right)$ is the initial value for the vorticity field.
We have \[
\hat{\omega}=\frac{d^{2}\phi}{dy^{2}}-k^{2}\phi\]
 From a mathematical point of view, we have to solve the equation
for $\phi$ \begin{equation}
\left(\frac{d^{2}}{dy^{2}}-k^{2}\right)\phi-\frac{U^{\prime\prime}}{U-c-i\epsilon}\phi=\frac{\omega\left(y,0\right)}{ik\left(U-c-i\epsilon\right)},\label{eq:Resolvent_Stream_Function}\end{equation}
 with the boundary condition for $\phi$ (here $\phi$ doubly periodic).
The solution of this boundary value problem $\phi\left[\omega(.,0)\right](y,c+i\epsilon)$
depends functionally on the initial value of the vorticity $\omega\left(y,0\right)$
($\phi$ is the resolvent operator for the stream function). This
resolvent operator encodes all the information about the temporal
evolution of the stream function and the vorticity field.\\

The homogeneous part of equation (\ref{eq:Resolvent_Stream_Function})
is the celebrated Rayleigh equation. It is also the equation for modes
($\psi=\phi\left(y\right)\exp\left[ik\left(x-\left(c+i\epsilon\right)t\right)\right]$)
of the linearized Euler equation (\ref{eq:Linearized_Euler}). It
reads\begin{equation}
\left(\frac{d^{2}}{dy^{2}}-k^{2}\right)\phi-\frac{U^{\prime\prime}}{U-c-i\epsilon}\phi=0,\label{eq:Rayleigh}\end{equation}
 with the flow boundary conditions. For neutral modes ($\epsilon=0$),
this is a non-classical boundary value problem, because of the possible
singularities associated to the vanishing of $U-c$. Any $y_{c}$
such that $U\left(y_{c}\right)=c$ is called a critical point for
the velocity $c$. For any $c$, the free motion on the streamline
$y=y_{c}$ is called the critical layer and has exactly the frequency
$kc$.

Any $y_{0}$ such that $U'(y_{0})=0$ (no shear, for instance for
velocity extrema) is called a stationary point of the jet profile,
corresponding to a stationary streamline. We then call $c_{_{0}}=U\left(y_{0}\right)$
a stationary velocity. If $y_{0}$ is a local extrema of $U$, we
note that when $y\rightarrow y_{0}$ (or equivalently $c\rightarrow c_{0}=U\left(y_{0}\right)$),
two critical layers, one on each side of the velocity extrema, merge
on a single one.

The range of the profile $U$ is the ensemble of velocities $c$ such
that $\min_{y}U\left(y\right)<c<\max_{y}U\left(y\right)$.\\

In the following we assume that the base flow is spectrally stable,
i.e., no unstable mode exist, which means that no solution to (\ref{eq:Rayleigh})
exist for any $c+i\epsilon$ with strictly positive $\epsilon>0$.
Then, as shown in section \ref{sub:resolvent}, equation (\ref{eq:Resolvent_Stream_Function})
has a unique solution for any $c+i\epsilon$ with strictly positive
$\epsilon>0$. We also assume that no neutral mode exist, which means
that no solutions to (\ref{eq:Rayleigh}) is found in the limit $\epsilon\rightarrow0^{+}$(a
more precise definition will be given bellow in terms of the dispersion
relation). This no mode assumption for equation (\ref{eq:Linearized_Euler_Fourier})
may seem strange, but it is indeed a generic situation when such flows
are stables. It is indeed a classical result that shear flow without
inflection points, or vortex with strictly decreasing vorticity profile
are stable and have no neutral mode \cite{Drazin_Reid_1981,Briggs_BDL_1970_Phys_Fluids}.
For instance, in section \ref{sec:Kolmogorov-flow}, we prove that
this hypothesis is also verified for the Kolmogorov flow $U\left(y\right)=\cos\left(y\right)$
as soon as the aspect ratio $\delta$ is larger than $1$. Actually
the only examples we are aware of, of stable flows for the 2D Euler dynamics, with neutral
modes that remain when we add small  perturbation to the flow, are cases with localized vorticity profile \cite{Schecter_Dubin_etc_Vortex_Crystals_2DEuler1999PhFl}.
Usually modes appear, when a parameter is changed, at the edge of
an instability ; four unstable (or two degenerate) unstable modes
eigenvalue then emerge from the continuous spectrum.

The case with neutral modes could be treated following the same lines
as what will be discussed bellow ; one would then have to separate
the contributions by the neutral modes from the contributions by the
continuous spectrum. The following discussion analyzes the contributions
by the continuous spectrum only.\\

\subsection{Large time asymptotic for the linearized Euler equation\label{sub:Large-time-asymptotic}}

In this section, we predict the large time asymptotic of the linearized
Euler equation, using Laplace transform tools. We prove results (\ref{eq:Vitesse_Asymptotics})
and (\ref{eq:Lungren}) and the mechanism of vorticity depletion at
stationary streamlines. The heart of the proof relies on the study
of the effect of critical layers, on the inhomogeneous Rayleigh equation
(\ref{eq:Resolvent_Stream_Function}), in the limit $\epsilon$ goes
to zero. This rather technical part is performed in section (\ref{sub:resolvent}).

The results of section (\ref{sub:resolvent}) are that the resolvent
streamfunction $\phi$, solution of (\ref{eq:Resolvent_Stream_Function}),
has a finite limit for small positive $\epsilon$ : \begin{equation}
\phi\left(y,c+i\epsilon\right)\underset{\epsilon\rightarrow0^{+}}{\rightarrow}\phi_{+}\left(y,c\right),\label{eq:limite_phi}\end{equation}
 even if singularities exist due to the critical layers. We prove
that for any $y$, $\phi_{+}\left(y,c\right)$ is twice differentiable
with respect to $c$, except for velocities $c$ that are in the interior
of the range of $U$. In this last case, for velocities $c$ that
are not stationary, $\phi_{+}$ is twice differentiable, except 
for $c=U\left(y\right)$. For $c=U\left(y\right)$, $\phi_{+}$ is continuous but not
differentiable there, and has a logarithmic singularity: for fixed
$y$, $\phi_{+}\left(y,c\right)=\Delta\phi_{c}\left(c-U\left(y\right)\right)\log\left(c-U\left(y\right)\right)+R\left(y,c\right)$,
where $R\left(y,.\right)$ is an analytic function of $c$. When $c=c_{0}$
is a stationary velocity, $\phi_{+}\left(y,c_{0}\right)$ is differentiable
with respect to $c$.

We think that all of the steps of these proofs could be easily made
rigorous from a mathematical point of view, by making explicit the
required hypothesis. An exception is for the limit of $\phi$ in the
case of critical layers for stationary points. We actually prove in
the following that a limit solution exist for $\epsilon=0$, but we
do not prove the convergence to it when $\epsilon\rightarrow0$. In
order to deal with this small gap in the proof, we will show, by numerically
computing $\phi\left(y,\epsilon\right)$, that this convergence actually
takes place.

These results (the limit and its properties) are the difficult aspects
of the discussion, from a mathematical point of view. Their technical
proof can be skipped at a first reading, the next sections can be
read independently by assuming these results. The discussion then
follows by performing the inverse Laplace transform and proving results
(\ref{eq:Vitesse_Asymptotics}) and (\ref{eq:Lungren}) in sections
\ref{sub:asymptotic-vorticity-field} and \ref{sub:asymptotic-velocity}.

\subsection{Limit for $\epsilon\rightarrow0^{+}$ of the resolvent operator \label{sub:resolvent}}

\subsubsection{The dispersion relation\label{sub:The-dispersion-relation}}

The equation defining the resolvent operator for the stream function
(\ref{eq:Resolvent_Stream_Function}) is of the type \begin{equation}
\frac{d^{2}\phi}{dy^{2}}+q(y)\phi=f(y)\label{eq:phi_inhomogeneous}\end{equation}
 with $q=-k^{2}-U''/(U-c-i\epsilon)$ and $f=\omega(y,0)/[ik(U-c-i\epsilon)]$.
This is a boundary value problem. In order to be precise, we treat
the case of a doubly periodic domain with the period $2\pi$, although
that is easily generalized to the case of a flow in a channel $y\in(a,b)$
with the boundary conditions $\phi(a)=\phi(b)=0$.

For $\epsilon\neq0$, the differential equation is not singular. We
consider the homogeneous equation \begin{equation}
\frac{d^{2}\phi}{dy^{2}}+q(y)\phi=0.\label{eq:phi_homogenous}\end{equation}
 We consider two independent solutions to (\ref{eq:phi_homogenous})
: $\phi_{1}$ is defined by $\phi_{1}\left(0\right)=1$ and $\phi'_{1}\left(0\right)=0$,
and $\phi_{2}$ is defined by $\phi_{2}\left(0\right)=0$ and $\phi'_{2}=1$
(here and bellow, primes are derivatives with respect to $y$). The
classical variation of the parameter computation then insures that a particular
solution to (\ref{eq:phi_inhomogeneous}) is \[
\phi_{p}\left(y\right)=-\left(\int_{0}^{y}\phi_{2}f\right)\phi_{1}\left(y\right)+\left(\int_{0}^{y}\phi_{1}f\right)\phi_{2}\left(y\right),\]
 and a general solution is \begin{equation}
\phi_{f}=\phi_{p}+a\phi_{1}+b\phi_{2},\label{eq:phi_f}\end{equation}
 where $a$ and $b$ are unknown constants. The necessary and sufficient
conditions for $\phi$ to be periodic are that $\phi(0)=\phi\left(2\pi\right)$
and $\phi'(0)=\phi'\left(2\pi\right)$. These conditions read \begin{equation}
M\left(\begin{array}{c}
a\\
b\end{array}\right)=\left(\begin{array}{c}
-\phi_{p}\left(2\pi\right)\\
-\phi'_{p}\left(2\pi\right)\end{array}\right)\,\,\,\mbox{with}\,\,\, M=\left(\begin{array}{cc}
\phi_{1}\left(2\pi\right)-1 & \phi_{2}\left(2\pi\right)\\
\phi'_{1}\left(2\pi\right) & \phi'_{2}\left(2\pi\right)-1\end{array}\right)\label{eq:a_b}\end{equation}
 This system has a single solution if and only if the determinant
of $M$ is nonzero, which gives the dispersion relation \begin{equation}
D\left(c+i\epsilon\right)\equiv\left[\phi_{1}\left(2\pi\right)-1\right]\left[\phi'_{2}\left(2\pi\right)-1\right]-\phi'_{1}\left(2\pi\right)\phi_{2}\left(2\pi\right)=0.\label{eq:dispersion}\end{equation}
 The existence of modes (nontrivial solutions to (\ref{eq:phi_homogenous}))
is then equivalent to the zero values of the dispersion relations.
When no mode exist, $D$ is nonzero and (\ref{eq:phi_inhomogeneous})
has thus a unique periodic solution $\phi_{f}$ (\ref{eq:phi_f}),
with $a$ and $b$ the unique solution to (\ref{eq:a_b}).\\

Turning back to the inhomogeneous Rayleigh equation, the preceding
discussion applies as soon as $\epsilon\neq0$. We assume that no
unstable mode exists, then $D\left(c+i\epsilon\right)$ is nonzero.
Then the inhomogeneous Rayleigh equation has a unique solution for
any $c+i\epsilon$, for nonzero $\epsilon$.

The limit $\epsilon\rightarrow0$ of $\phi_{f}\left(c+i\epsilon\right)$
is nontrivial due to the existence of critical layers $y_{c}\left(c\right)$,
for which the Rayleigh equation becomes singular. We study this limit
in the following sections.

\subsubsection{Limit $\epsilon\rightarrow0^{+}$ for isolated critical layers \label{sub:Single-Critical-Layer}}

We consider fixed values of $c$ which are on the range of $U$:
$\min_{y}\left\{ U\left(y\right)\right\} <c<\max_{y}\left\{ U\left(y\right)\right\} $.
In such a case, for any value of $c$, there exist one or several
points $y_{l}$ such that $U(y_{l})=c$. The inhomogeneous Rayleigh
equation is then singular at such critical layers. We discuss in this
section the case $U'(y_{l})\neq0$ (isolated critical layers). The
case $U'(y_{l})=0$ will be treated in the next section.

In order to properly study the limit $\epsilon\rightarrow0$, we first
build a solution to the homogeneous equation (\ref{eq:Rayleigh}),
which is regular at one of the critical layers $y=y_{l}$. We define
$\phi_{r}\left(y,c\right)$ as the solution to (\ref{eq:Rayleigh})
with $\phi_{r}\left(y_{l},c\right)=0$ and $\phi_{r}^{'}\left(y_{l},c\right)=1$.
From (\ref{eq:Rayleigh}), we have $\phi_{r}^{''}(y_{l},c)=U''(y_{l})/U'(y_{c})$.
We then have the expansion \begin{equation}
\phi_{r}\left(y,c\right)=(y-y_{l}\left(c\right))\left[1+\frac{U''\left(y_{l}\left(c\right)\right)}{2U'\left(y_{l}\left(c\right)\right)}\left(y-y_{l}\left(c\right)\right)+o\left(y-y_{l}\right)\right]\label{eq:DLpsi1}\end{equation}
 It can be shown that the solution $\phi_{r}\left(y,c\right)$ is
an analytic function of $y$ in the vicinity of $y_{l}$, if we suppose
that $U\left(y\right)$ is analytic in a vicinity of $y_{l}$. Moreover,
from the definition $U\left(y_{l}\left(c\right)\right)=c$, because
$U'\left(y_{l}\right)\neq0$, then $y_{l}\left(c\right)$ is analytic
in a vicinity of $c$ and $dy_{l}/dc=1/U'\left(y_{l}\right)$. The
solution $\phi_{r}\left(y,c'\right)$ has then an analytic continuation
for complex $c'$ in the vicinity of $c$. \\

A classical result of the theory of differential equation of second
order is that, if we already know a solution $\phi_{r}$, all other
solutions $\phi$ are expressed in terms of $\phi_{r}$ by quadratures.
 The recipe for this is to look for solutions under the form $\phi=u\phi_{r}$,
look for the equation verified by $u$ and integrate it. We apply
this recipe to the inhomogeneous Rayleigh equation (\ref{eq:Resolvent_Stream_Function}).
Then any solution $\phi$ to (\ref{eq:Resolvent_Stream_Function})
is expressed as \begin{equation}
\phi(y)=d\phi_{r}(y)+\phi_{r}(y)\int_{y_{0}}^{y}\frac{(e+f)}{\phi_{r}^{2}}\,\,\,\mbox{\,\,\, with}\label{eq:phi_integrale}\end{equation}
 \[
f(y)=\int_{y_{0}}^{y}dy_{2}\,\frac{\omega\left(y_{2},0\right)\phi_{r}(y_{2})}{ik\left(U(y_{2})-c-i\epsilon\right)},\]
 and where $d$ and $e$ are constants.

We study the behavior of the previous expression close to $y_{l}$.
We first note that $f$ is analytic close to $y_{l}$. Then using
the expansion (\ref{eq:DLpsi1}), we conclude that \begin{equation}
\phi(y,c+i\epsilon)=d\phi_{r}(y,c+i\epsilon)+g\phi_{r}(y,c+i\epsilon)\log\left(y-y_{l}(c+i\epsilon)\right)+e\phi_{g}\left(y,c+i\epsilon\right)+\phi_{h}\left(y,c+i\epsilon\right)\label{eq:phi_h}\end{equation}
 where $\phi_{g}$ and $\phi_{h}$ are analytic functions of $y$
close to $y_{l}$, and where $g$ is a constant that depends on $f(y_{l})$,
$f'\left(y_{l}\right)$ and $e$.

The interpretation of (\ref{eq:phi_h}) depends on which determination
of the logarithm we use. Using $dy_{l}/dc=1/U'\left(y_{l}\right)$
(discussed above), we have $y_{l}\left(c+i\epsilon\right)=y_{l}\left(c\right)+i\epsilon/U'\left(y_{l}\right)+o\left(\epsilon\right)$.
We choose a determination of the logarithm such that $\log\left(y-y_{l}(c)-i\epsilon/U'\left(y_{l}\right)\right)$
remains analytic for positive $\epsilon$. Then the study of the limit
$\epsilon\rightarrow0$ of equation (\ref{eq:phi_f}) is easily done,
we denote this limit $\phi\left(y,c+i0\right)$. Using that $\phi_{r}$
and $\phi_{g}$ depends analytically on $c$, we obtain \begin{equation}
\phi(y,c+i0)=d\phi_{r}(y,c)+g\phi_{r}(y,c)\log\left|y-y_{l}\right|+e\phi_{g}\left(y,c\right)+\phi_{h}\left(y,c\right)\,\,\,\mbox{for}\,\,\, y>y_{l}\,\,\,\mbox{and}\label{eq:determination_logarithm1}\end{equation}
 \begin{equation}
\phi(y,c+i0)=(d-i\pi\mbox{sgn}\left(U'(y_{l})\right)g)\phi_{r}(y,c)+g\phi_{r}(y,c)\log\left|y-y_{l}\right|+e\phi_{g}\left(y,c\right)+\phi_{h}\left(y,c\right)\,\,\,\mbox{for}\,\,\, y<y_{l}\label{eq:determination_logarithm_2}\end{equation}
 where $\mbox{sgn}\left(U'(y_{l})\right)$ is the sign of $U'(y_{l})$.

From this, we conclude that, for given $d$ and $e$, the solution
to the inhomogeneous Rayleigh equation (\ref{eq:Resolvent_Stream_Function})
converges, for $\epsilon\rightarrow0$, towards a function $\phi\left(y,c+i0\right)$,
which is an analytical function of $y$, except for $y=y_{l}$ where
it has a logarithmic singularity. It is continuous at $y=y_{l}$.
This result is valid close to any single critical layer $y_{l}$. If two or several
critical layer $y_{l,i}$ exist in the interval $y\in(0,2\pi)$, then
the result is easily extended, and holds with a singularity at each
critical layer. \\

Is this result also true for the solution $\phi_{\omega}(y,c+i\epsilon)$
of the inhomogeneous Rayleigh equation with boundary conditions ?
In order to answer this, we now turn again to the construction of
section \ref{sub:The-dispersion-relation}. The result of the previous
paragraph is applied alternatively to $\phi_{1}(y,c+i\epsilon)$,
$\phi_{2}\left(y,c+i\epsilon\right)$ and to $\phi_{p}(y,c+i\epsilon)$.
We thus conclude that all these three functions have well-defined
limits for $\epsilon\rightarrow0^{+}$, and that these limits are
continuous functions of $y$, which have logarithmic singularities
in their derivative for each critical layer. We can then extent the
definition of the dispersion relation to $\epsilon\rightarrow0^{+}$
with $D_{+}\left(c\right)=\lim_{\epsilon\rightarrow0^{+}}D\left(c+i\epsilon\right)$.
\textbf{$D_{+}\left(c\right)$} verifies (\ref{eq:dispersion}) for
which we have proved that all terms have a finite limit when $\epsilon\rightarrow0^{+}$.
Then we conclude that the two parameters $a$ and $b$ in equation
(\ref{eq:phi_f}) have finite limits when $\epsilon\rightarrow0^{+}$.
These limit values are given by equation (\ref{eq:a_b}), where each
term has a finite limit. \\

We thus conclude that the solution to the Rayleigh equation with boundary
conditions $\phi_{\omega}(y,c+i\epsilon)$ has a finite limit $\phi_{+}\left(y,c\right)$
for $\epsilon\rightarrow0^{+}$. Moreover, $\phi_{+}\left(y,c\right)$
is a continuous function of $y$ that has a logarithmic singularity
at each critical layer, giving a finite jump for the first derivative.

Let us denote $\Delta\phi_{+}$ this jump. From the previous analysis
we know that\begin{equation}
\phi_{+}\left(y,c\right)=a+b\left(y-y_{l}\right)+\Delta\phi_{+}(y-y_{l})\log\left|y-y_{l}\right|+\mathcal{O}\left(y-y_{l}\right)^{2}\label{eq:phi_plus}\end{equation}
 Using this expansion, a direct analysis of the dominant term in (\ref{eq:Resolvent_Stream_Function})
leads to \[
\Delta\phi_{+}=\frac{\omega\left(y_{l},0\right)+ikU''\left(y_{l}\right)\phi_{+}\left(y_{l},c\right)}{ikU'\left(y_{l}\right)}.\]
 The jump in the derivative thus depends on the value of $\phi_{+}$
which is a non local quantity ($\phi_{+}$ depends on the whole profile
$U$).

Because $\phi_{r}$ and $y_{l}$ are analytic functions of $c$, the
construction of $\phi_{+}$ can be extended analytically when $c$
is varied. Then from (\ref{eq:phi_plus}), using $y_{l}\left(c'\right)=y_{l}\left(c\right)+(c'-c)/U'(y_{l})$,
one sees that, for fixed $y$:\begin{equation}
\phi_{+}\left(y,c\right)=\Delta\phi_{c}\left(U\left(y\right)-c\right)\log\left(U\left(y\right)-c\right)+\phi_{a}\left(y,c\right),\label{eq:singularite_c}\end{equation}
 where $\phi_{a}\left(y,c\right)$ is analytic close to $c=U\left(y\right)$
and with \begin{equation}
\Delta\phi_{c}=\frac{\Delta\phi_{+}}{U'\left(y\right)}=\frac{\omega\left(y,0\right)+ikU''\left(y\right)\phi_{+}\left(y,U\left(y\right)\right)}{ik\left(U'\left(y\right)\right)^{2}}.\label{eq:Delta_Phi_c}\end{equation}
 \\

We illustrate the preceding results with numerical solutions of the
inhomogeneous Rayleigh equations on a doubly periodic domain, and
the base flow $U\left(y\right)=\cos{y}$ (the Kolmogorov flow).

We follow the algorithm described in section \ref{sub:The-dispersion-relation},
that is, computing $\phi_{1}$, $\phi_{2}$, and $\phi_{p}$, and
then by using them, computing the solution to the inhomogeneous Rayleigh
equation for $c'=c+i\epsilon$ for small but nonzero value of $\epsilon$.
In order to numerically compute the solution to the differential equations
(for $\phi_{1}$, $\phi_{2}$, and $\phi_{p}$), we use an adaptive
method to deal with the singularity close to the critical layers.
An extreme precision is required in order to obtain satisfactory results.

In order to test the quality of the numerical simulations, we compute
the Wronskian $W=\phi_{1}\left(y\right)\phi'_{2}\left(y\right)-\phi_{2}\left(y\right)\phi'_{1}\left(y\right)$.
From the general theory of differential equations of second order,
we know that $W$ does not depend on $y$. Here, from the values of
$\phi_{1}$ and $\phi_{2}$ at $y=0$, given by their definition,
we have $W=1$. We test the accuracy of this in all our numerical
simulations. For instance in the case of simple critical layers, using
the Matlab function \emph{ode45,} and fixing the relative error and
the absolute error parameters of this function to $10^{-13}$, we
obtain solutions for which errors in $W$ are typically smaller than
$10^{-6}$ for $\epsilon=10^{-4}$.

\begin{figure}
\begin{centering}
\includegraphics[width=0.75\columnwidth]{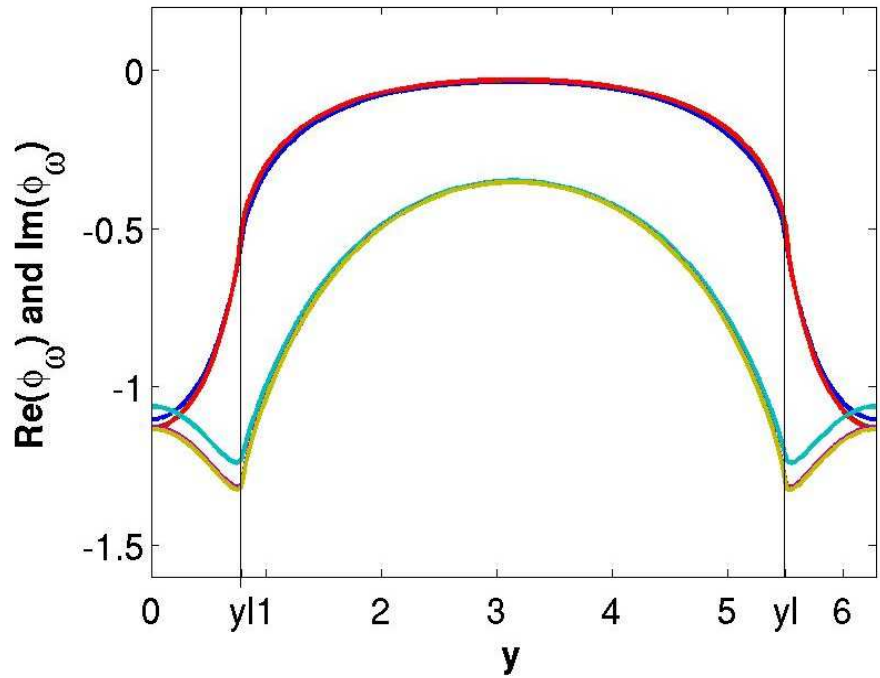}
\par\end{centering}

\caption{\label{fig:phi_omega_pi_4} The real ($\Re$, upper set of curves)
and the imaginary part ($\Im$, lower set of curves) for the solution
$\phi_{\omega}\left(y,c+i\epsilon\right)$ to the inhomogeneous Rayleigh
equation (\ref{eq:Resolvent_Stream_Function}), in the case $\omega(y,0)=1$,
$c=\sqrt{2}/2$ ($y_{l}=\pi/4$ and $y_{l}=7\pi/4$), $k=1.5$. The
different curves show the results for $\epsilon=10^{-2}$(blue ($\Re$)
and light blue ($\Im$)), $\epsilon=10^{-3}$ (green ($\Re$) and
magenta ($\Im$)) and $\epsilon=10^{-4}$ (red ($\Re$) and yellow
($\Im$)). The curves for $\epsilon=10^{-3}$ and $\epsilon=10^{-4}$
are nearly indistinguishable, showing good convergence. }

\end{figure}

\begin{figure}
\begin{centering}
\includegraphics[width=0.75\columnwidth]{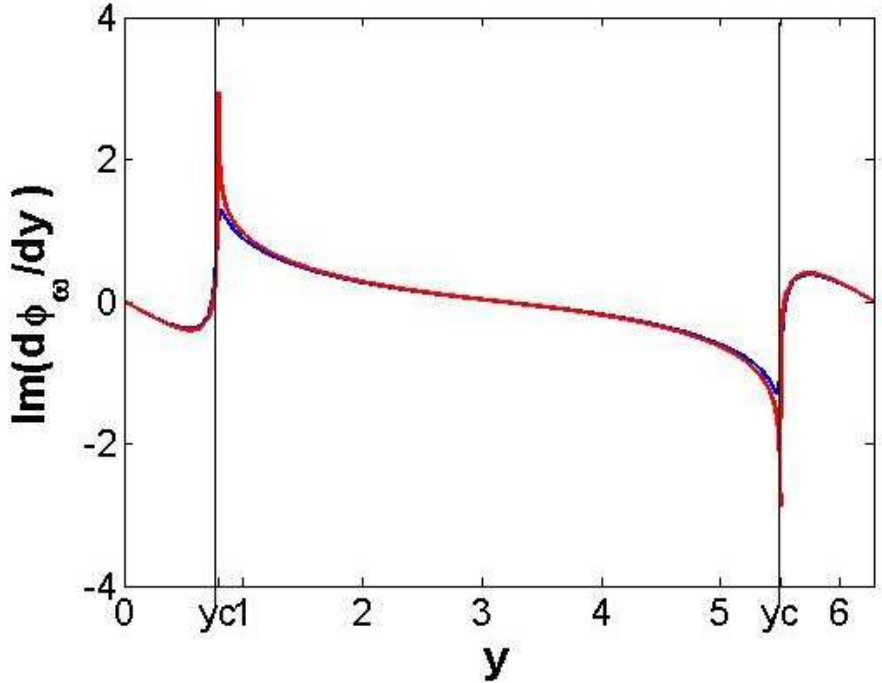}
\par\end{centering}

\caption{\label{fig:phi_omega_prime_pi_4} Same as the previous figure, but
for the imaginary part of the derivative $\phi'_{\omega}\left(y,c+i\epsilon\right)$
of the solution to the inhomogeneous Rayleigh equation (\ref{eq:Resolvent_Stream_Function})
(blue for $\epsilon=10^{-2}$, green for $\epsilon=10^{-3})$ and red
for $\epsilon=10^{-4}$. The green and red curves are nearly indistinguishable.}

\end{figure}

Figure \ref{fig:phi_omega_pi_4} shows the real and the imaginary
part for the solution $\phi_{\omega}\left(y,c+i\epsilon\right)$ to
the inhomogeneous Rayleigh equation (\ref{eq:Resolvent_Stream_Function}),
in the case $\omega(y,0)=1$, $c=\sqrt{2}/2$ ($y_{l}=\pi/4$ and
$y_{l}=7\pi/4$), $k=1.5$ for the values $\epsilon=10^{-2}$, $\epsilon=10^{-3}$
and $\epsilon=10^{-4}$. This illustrates the convergence of the solutions
$\phi_{\omega}\left(y,c+i\epsilon\right)$ to a continuous function
$\phi_{\omega}\left(y,c+i0\right)$. The visible kinks close to the
critical layers suggest the discontinuity of the derivative. This
is actually verified and illustrated in figure \ref{fig:phi_omega_prime_pi_4},
that shows the derivative with respect to $y$, $\phi'_{\omega}\left(y,c+i\epsilon\right)$.

\subsubsection{Limit $\epsilon\rightarrow0^{+}$ for critical layer of stationary
streamlines \label{sub:Limit-Stationary-points} }

We now consider the case of a critical layer that correspond to a
stationary streamline ($y_{l}=y_{0}$, where $U(y_{0})=c_{0}$, $U'(y_{0})=0$
and $U''\left(y_{0}\right)\neq0$).\\

In order to properly study the limit $\epsilon\rightarrow0$ in this
case, we first build a solution to the homogeneous equation (\ref{eq:Rayleigh}),
which is regular at the critical layer $y=y_{0}$. We define $\phi_{r}\left(y,c_{0}\right)$
as the solution to (\ref{eq:Rayleigh}) with $\phi_{r}\left(y_{0},c_{0}\right)=0$,
$\phi_{r}^{'}\left(y_{0},c_{0}\right)=0$ and $\phi_{r}^{''}\left(y_{0},c_{0}\right)=1$.
Such a solution can be shown to exist by a series expansion in powers
of $(y-y_{0})$. It can be shown that $\phi_{r}\left(y,c_{0}\right)$
is an analytic function of $y$ in the vicinity of $y_{0}$, if we
suppose that $U\left(y\right)$ is analytic in the vicinity of $y_{0}$.

However, by contrast to the case of isolated critical layers analyzed
in section (\ref{sub:Single-Critical-Layer}), the solution $\phi_{r}\left(y,c_{0}\right)$
is not analytic in the vicinity of $c_{0}$. Then the approach of the previous section can not be generalized. 

Let us first prove that it exists a solution to the inhomogeneous
Rayleigh equation, for $\epsilon=0$, which is continuously differentiable
at $y_{0}$. We start from expression (\ref{eq:phi_integrale}). $f$
is analytic in $y_{0}$. Let us choose $e=-f(y_{0})$ and $b=0$.
Then we obtain the particular solution \begin{equation}
\phi_{\omega,0}(y)=\phi_{r}(y)\int_{y_{0}}^{y}\frac{(f-f\left(y_{0}\right))}{\phi_{r}^{2}}.\label{eq:phi_omega_0}\end{equation}
 Noting that the expansion of $\phi_{r}$ begins at order 2 in $(y-y_{0})$,
we easily prove that \begin{equation}
\phi_{\omega,0}(y)=\phi_{i}\left(y\right)+g\phi_{r}\left(y\right)\log\left|y-y_{0}\right|,\label{eq:phi_omega_0_log}\end{equation}
 where $g$ is a constant and $\phi_{i}$ is an analytic function
of $y$. We note that $\phi_{\omega,0}(y)$ is continuously differentiable
at the critical layer $y_{0}$. This solution is defined locally,
in an interval where $\phi_{r}$ has no other zero than $y_{0}$.
However, it can be extended to the whole interval $y\in(0,2\pi)$,
because equation (\ref{eq:Resolvent_Stream_Function}) is not singular
in other points than $y_{0}$.

By contrast to the situation obtained for isolated critical points,
we can not make an analytical continuation of the solution (\ref{eq:phi_omega_0})
for complex $c_{0}+i\epsilon$. We thus follow another route.

We note that we can add $b\phi_{i}$ to $\phi_{\omega,0}(y)$, where
two different values for $b$ can be chosen for $y<y_{0}$ and for
$y>y_{0}$. The function \begin{equation}
\left\{ \begin{array}{c}
\phi_{+}(y,c_{0})=b^{-}\phi_{r}(y)+\phi_{\omega,0}(y)\,\,\,\mbox{for}\,\,\, y<y_{0}\\
\phi_{+}(y,c_{0})=b^{^{+}}\phi_{r}(y)+\phi_{\omega,0}(y)\,\,\,\mbox{for}\,\,\, y>y_{0}\end{array}\right.\label{eq:phi_omega_s}\end{equation}
 is actually a solution to the inhomogeneous Rayleigh equation (\ref{eq:Resolvent_Stream_Function})
for any $y\neq y_{0}$ which is continuously differentiable in $y_{l}$.
It is thus a solution to (\ref{eq:Resolvent_Stream_Function}).

The values of $b^{+}$ and $b^{-}$ can be chosen in order to satisfy
the boundary conditions. For instance, for $2\pi$-periodic solutions,
the boundary conditions are equivalent to \begin{equation}
N\left(\begin{array}{c}
b^{-}\\
b^{+}\end{array}\right)=\left(\begin{array}{c}
\phi_{\omega,0}\left(2\pi\right)-\phi_{\omega,0}\left(0\right)\\
\phi'_{\omega,0}\left(2\pi\right)-\phi'_{\omega,0}\left(0\right)\end{array}\right)\,\,\,\mbox{with}\,\,\, N=\left(\begin{array}{cc}
\phi_{r}\left(0\right) & -\phi_{r}\left(2\pi\right)\\
\phi'_{r}\left(0\right) & -\phi'_{r}\left(2\pi\right)\end{array}\right)\label{eq:N}\end{equation}
 The determinant of $N$ then plays the role of a dispersion relation
for neutral mode associated to the stationary streamlines. It reads
\begin{equation}
D_{s}=-\phi_{r}\left(0\right)\phi'_{r}\left(2\pi\right)+\phi_{r}\left(2\pi\right)\phi'_{r}\left(0\right)\label{eq:dispersion_S}\end{equation}
 When no such mode exists, equation (\ref{eq:N}) is solved, and we
obtain a solution to the inhomogeneous Rayleigh equation that verifies
the boundary conditions.

We have thus constructed a solution to the inhomogeneous Rayleigh
equation for real $c_{0}=U\left(y_{0}\right)$, where $y_{0}$ is
a stationary point of $U$ such that $U''\left(y_{0}\right)\neq0$.
\\

We illustrate the preceding results with numerical solutions of the
inhomogeneous Rayleigh equations on a doubly periodic domain, for
the Kolmogorov base flow $U\left(y\right)=\cos{y}$.

The numerical computation follows the same rules are the one described
in section \ref{sub:Single-Critical-Layer}. We note that using the
Matlab function \emph{ode45,} and fixing the relative error and the
absolute error parameters of this function to $10^{-13}$, we obtain
solutions for which errors in $W$ are typically smaller than $10^{-2}$
for $\epsilon=10^{-3}$. It is thus much harder to obtain good quality
numerical simulation in that case, than in the case of isolated critical
layers discussed in section \ref{sub:Single-Critical-Layer}.

\begin{figure}
\begin{centering}
\includegraphics[width=0.75\columnwidth]{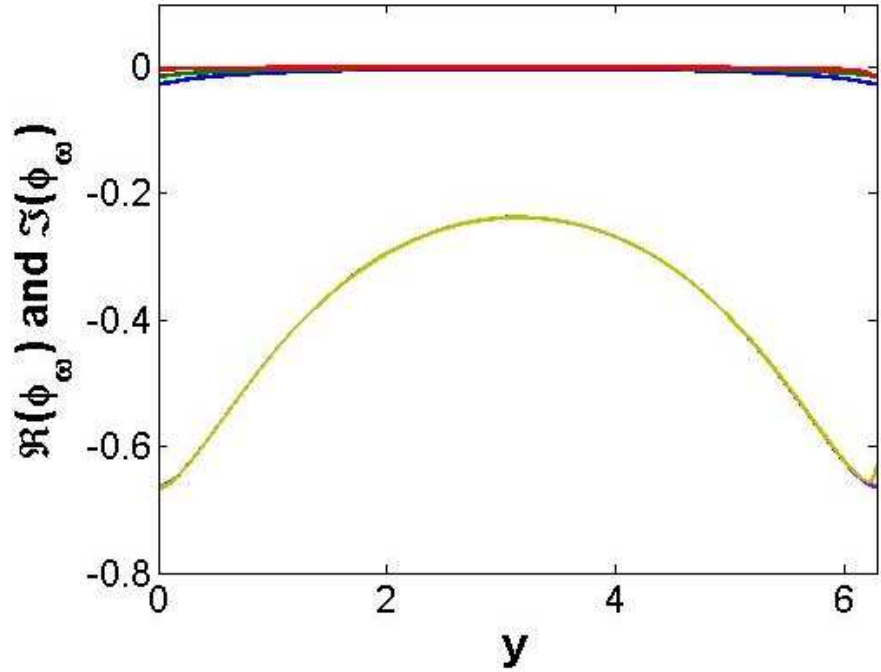}
\par\end{centering}

\caption{\label{fig:phi_omega_c1} The real ($\Re$) and the imaginary ($\Im$)
part for the solution $\phi_{\omega}\left(y,c_{0}+i\epsilon\right)$
to the inhomogeneous Rayleigh equation (\ref{eq:Resolvent_Stream_Function}),
in the case $\omega(y,0)=1$; with a critical layer corresponding
to a stationary point ($c_{0}=1$, $y_{l}=0$) and with $k=1.5$.
The different curves show the results for $\epsilon=10^{-2}$ (blue
($\Re$) and light blue ($\Im$)), $\epsilon=5.10^{-3}$ (green ($\Re$)
and magenta ($\Im$)) and $\epsilon=10^{-3}$ (red ($\Re$) and yellow
($\Im$)). The curves for $\epsilon=5.10^{-3}$ and $\epsilon=10^{-3}$
are indistinguishable, showing good convergence. The real part is
the ensemble of curves that converge to zero. }

\end{figure}

\begin{figure}
\begin{centering}
\includegraphics[width=0.75\columnwidth]{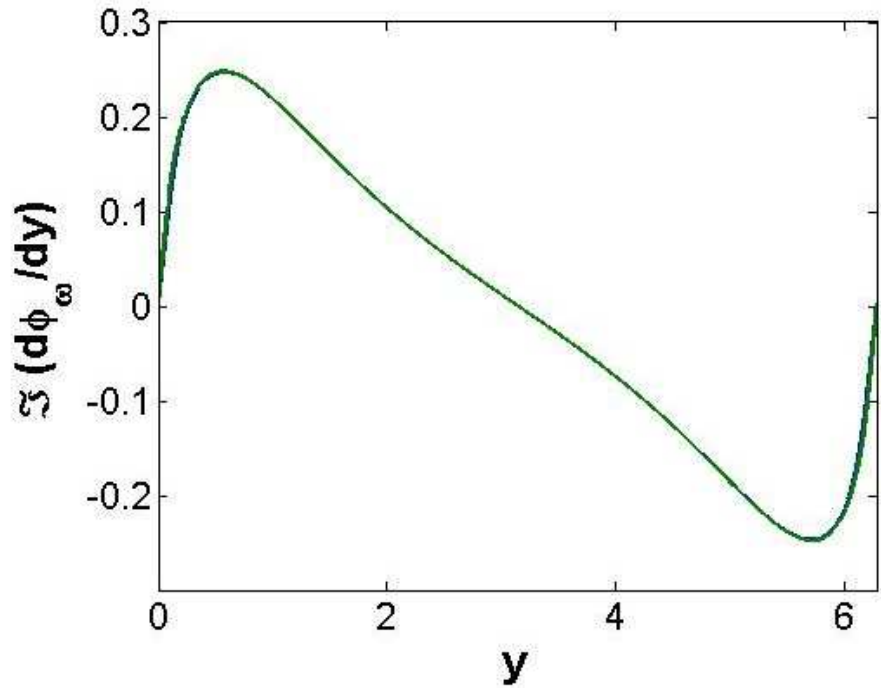}
\par\end{centering}

\caption{\label{fig:phi_omega_prime_c1} Same as the previous figure, but for
the imaginary part of the derivative $\phi'_{\omega}\left(y,c_{0}+i\epsilon\right)$
of the solution to the inhomogeneous Rayleigh equation (\ref{eq:Resolvent_Stream_Function}).
All three curves are indistinguishable.}

\end{figure}

Figure \ref{fig:phi_omega_c1} shows the real and the imaginary part
for the solution $\phi_{\omega}\left(y,c_{0}+i\epsilon\right)$ to
the inhomogeneous Rayleigh equation (\ref{eq:Resolvent_Stream_Function}),
in the case $\omega(y,0)=1$; with a critical layer corresponding
to a stationary point ($c_{0}=1$, $y_{0}=0$) and with $k=1.5$ ;
for the values $\epsilon=10^{-2}$, $\epsilon=5.10^{-3}$ and $\epsilon=10^{-3}$.
This illustrates the convergence of the solutions $\phi_{\omega}\left(y,c_{0}+i\epsilon\right)$
to a continuous function $\phi_{+}\left(y,c_{0}\right)$. It turns
out that the real part converges to zero. The same results are also
presented for the derivative with respect to $y$, $\phi'_{\omega}\left(y,c_{0}+i\epsilon\right)$,
on figure \ref{fig:phi_omega_prime_c1}.\\

We now turn to the derivation of a property of such solutions, that
will be very important in the discussion of the asymptotic behavior
of the linearized 2D Euler equations. We have shown that $\phi_{+}\left(y,c_{0}\right)$
is continuously differentiable at $y_{l}$, and has a second order
logarithmic singularity at $y_{0}$ (see (\ref{eq:phi_omega_0_log})
and (\ref{eq:phi_omega_s})). Then a direct inspection of the leading
singularity in equation (\ref{eq:Resolvent_Stream_Function}), of
order $(y-y_{c})^{-1}$, leads to the conclusion that\begin{equation}
ikU''\left(y_{0}\right)\phi_{+}\left(y_{0},U\left(y_{0}\right)\right)+\omega(y_{0},0)=0\label{eq:phi_omega_yc}\end{equation}

\subsubsection{The asymptotic vorticity field\label{sub:asymptotic-vorticity-field}}

Using the results of the previous section, we prove in this section
that the vorticity field converges, for large time, towards a field
oscillating at a multiple of the streamline frequency. More precisely,
we prove that \begin{equation}
\omega\left(y,t\right)\underset{t\rightarrow\infty}{\sim}\omega_{\infty}\left(y\right)\exp\left(-ikU(y)t\right)+\mathcal{O}\left(\frac{1}{t^{\gamma}}\right)\label{eq:omega_asymptotic}\end{equation}
 In particular, for any stationary point $y_{0}$, $\omega_{\infty}\left(y_{0}\right)=0$.
This essential property means that the vorticity cancels rapidly at
any stationary streamline. This is the mechanism of vorticity depletion
at the stationary streamlines, discussed in the introduction.\\

Using (\ref{eq:limite_phi}) and (\ref{eq:Resolvent_vorticity}) we
have \begin{equation}
\hat{\omega}(y,c+i\epsilon)\underset{\epsilon\rightarrow0^{+}}{\sim}\frac{ikU''\left(y\right)\phi_{+}\left(y,c\right)+\omega\left(y,0\right)}{ik\left(U(y)-c-i0^{+}\right)}\label{eq:resolvent_singularity}\end{equation}
Thanks to the analysis of the properties of $\phi$ in the previous section, we know that all its singularities are integrable. We thus see that for any $c$, $\hat{\omega}(.,c)$ has non integrable singularities
at each critical layer $y_{l}$. For fixed $y$, $\hat{\omega}(y,.)$
has a single singularity for the velocity $c=U(y)$. Using (\ref{eq:resolvent_singularity}),
we write the inverse Laplace transform (\ref{eq:inverse_Laplace})
on the Bromwich contour defined by $p=-ik(c+i\epsilon)$ with $\epsilon>0$
and $-\infty\leq c\leq+\infty.$ \begin{equation}
\omega\left(y,t\right)=\frac{1}{2\pi i}\int_{-\infty}^{+\infty}dc\,\frac{\exp\left(-ik(c+i\epsilon)t\right)}{U(y)-c-i\epsilon}[ikU''\left(y\right)\phi\left(y,c+i\epsilon\right)+\omega(0,y)]\label{eq:vorticity_inverse_Laplace}\end{equation}
 We first estimate the contribution of the pole \[
\frac{1}{2\pi i}\int_{-\infty}^{+\infty}dc\,\frac{\exp\left(-ik(c+i\epsilon)t\right)}{U(y)-c-i\epsilon}[ikU''\left(y\right)\phi_{+}\left(y,U(y)\right)+\omega(0,y)]=\left[ikU''\left(y\right)\phi_{+}\left(y,U(y)\right)+\omega\left(y,0\right)\right]\exp\left(-ikU(y)t\right),\]
 using the standard deformation of the contour of the complex integral,
and the residue theorem. The remainder contribution to the vorticity
(\ref{eq:vorticity_inverse_Laplace}) is then \[
\frac{kU''\left(y\right)}{2\pi}\int_{-\infty}^{+\infty}dc\,\frac{\phi_{+}\left(y,c\right)-\phi_{+}\left(y,U(y)\right)}{U(y)-c}\exp\left(-ikct\right).\]
 This integral is an oscillating integral. For large times, it thus
gives a contribution of order $\mathcal{O}\left(1/t^{\gamma}\right)$
where $\gamma$ depends on the order differentiability of $\phi_{+}\left(y,c\right)$
as a function of $c$. The result (\ref{eq:omega_asymptotic}) is
thus proved, and we have \begin{equation}
\omega_{\infty}\left(y\right)=ikU''\left(y\right)\phi_{+}\left(y,U(y)\right)+\omega\left(y,0\right).\label{eq:omega_asymptotic_phi}\end{equation}

We remark that for any point $y_{1}$ where $U''(y_{1})$
vanishes, $\omega_{\infty}(y_{1})=\omega(y_{1},0)$. This could have
been anticipated as for such points $y_{1}$, from (\ref{eq:Linearized_Euler_Fourier}),
we trivially have $\omega\left(y_{1},t\right)=\omega(y_{1},0)\exp\left(-ikU\left(y_{1}\right)t\right)$
for any time $t$. \\

Using (\ref{eq:phi_omega_yc}) and (\ref{eq:omega_asymptotic_phi}),
we deduce that \[
\omega_{\infty}\left(y_{0}\right)=0.\]
 This result means that the vorticity tends to zero for large time
for any stationary streamlines $y_{0}$. This vorticity depletion
at the stationary streamline is, from a mathematical point of view,
a nontrivial consequence of the Laplace transform analysis, and of
the regularization of the resolvent operator at stationary velocities.
As will be illustrated in section \ref{sec:Kolmogorov-flow}, using
numerical simulation, it is a striking dynamical effect leading to
the disappearance of any filament in the area close to the critical
layer of a stationary point of the profile $U$. This has a large
qualitative impact on the flow structure and evolution.

This effect comes from the term $v_{y}U''\left(y\right)$ in the linearized
Euler equation (\ref{eq:Linearized_Euler_Fourier}) ; it is thus a
consequence of the effect of the transverse velocity on the background
vorticity. Because $v_{y}$ is a non local quantity, depending on
the evolution of the vorticity field everywhere in the domain, this
effect is a non local, non trivial one that we are not able to explain
easily heuristically. \\

Besides these theoretical results, the Laplace tools are very interesting
as they allow the computation of asymptotic behavior of the flow without
relying on a complex direct numerical computation. Moreover, whereas
in asymptotic approaches like the Lundgren's one, where the asymptotic profile
$\omega_{\infty}\left(y\right)$ is not determined, here we can compute
it from (\ref{eq:omega_asymptotic_phi}).

Using this last procedure and the numerical computations of the resolvent
$\phi_{\omega}$, described in sections \ref{sub:Single-Critical-Layer}
and \ref{sub:Limit-Stationary-points}, we compute the asymptotic vorticity
profiles. They are represented in figures \ref{fig:w1_abs_fin_num_theor},
\ref{fig:w1_abs_fin_vs_aratio} and \ref{fig:w1_abs_fin_num_theor_i1} ;
and discussed in more details in section \ref{sub:numerical-computation}.

\subsubsection{The asymptotic velocity field\label{sub:asymptotic-velocity}}

In this section we study the asymptotic behavior of the velocity field.
We prove that the velocity field decays algebraically for large times: \begin{equation}
v_{x}(y,t)\underset{t\rightarrow\infty}{\sim}\frac{\omega_{\infty}\left(y\right)}{ikU'(y)}\frac{\exp\left(-ikU(y)t\right)}{t}\,\,\,\mbox{and}\label{eq:Vitesse-x-algebraic}\end{equation}
 \begin{equation}
v_{y}(y,t)\underset{t\rightarrow\infty}{\sim}\frac{\omega_{\infty}\left(y\right)}{ik\left(U'(y)\right)^{2}}\frac{\exp\left(-ikU(y)t\right)}{t^{2}};\label{eq:Vitesse-y-algebraic}\end{equation}
 where $\omega_{\infty}$ is the asymptotic vorticity profile (\ref{eq:omega_asymptotic},\ref{eq:omega_asymptotic_phi}).

We first explain this result starting from the asymptotic vorticity
derived in the previous section (\ref{eq:omega_asymptotic}), and
using large time asymptotic behavior of oscillating integrals. This
argument is heuristically very interesting. However it will be valid
only when the contributions of stationary points $y_{0}$ are negligible,
and when the convergence of the vorticity towards the asymptotic vorticity
is sufficiently rapid. Indeed, the convergence towards the Lundgren's
ansatz has to be uniformly more rapid than the derived algebraic laws.
This last point is actually true only for strictly monotonic velocity
profiles $U$ as will be discussed below.

In order to give a proof of the results (\ref{eq:Vitesse-x-algebraic})
and (\ref{eq:Vitesse-x-algebraic}) valid also for profile $U$
with stationary streamlines, we give a more general argument based
on Laplace transform at the end of this section.

\paragraph{Oscillating integrals}

We begin with the expression of the velocity from the vorticity field
using a Green function formalism. We have

\begin{equation}
\mathbf{v}(y,t)=\int dy'\,\mathbf{G}_{k}(y,y')\omega(y',t),\label{eq:Evolution_velocity}\end{equation}
where $\mathbf{G}_{k}$ is defined from (\ref{eq:relation_omega_v}):
\[
\mathbf{G}_{k}(y,y')=\left(-\frac{\partial H_{k}}{\partial y},ikH_{k}\right)\left(y,y'\right)\,\,\,\mbox{with}\,\,\,\frac{\partial^{2}H_{k}}{\partial y^{2}}-k^{2}H_{k}=\delta\left(y-y'\right),\]
 with the periodic boundary conditions on $y$. Using the asymptotic
result on the vorticity field, we thus have \begin{equation}
\mathbf{v}(y,t)\underset{t\rightarrow\infty}{\sim}\int dy'\,\mathbf{G}_{k}\left(y,y'\right)\omega_{\infty}(y')\exp\left(-ikU(y)t\right).\label{eq:velocity_ocillating_integral}\end{equation}

We consider the asymptotic behavior, for large times $t$, of the
oscillating integral (\ref{eq:velocity_ocillating_integral}). Since
Kelvin, very classical results do exist for the asymptotic behavior
of such integrals, the most famous result being the stationary
phase approximation. Such results are discussed in appendix A.

An essential point, which makes this case different from the more classical
ones, is that the Green function $\mathbf{G}_{k}\left(y,y'\right)$
is not smooth everywhere: it is smooth except for the singularity
when $y=y'$. We prove in appendix A that \emph{if $U\left(y\right)$
has no stationary points} (for all $y$, $U'\left(y\right)\neq0$)\textbf{,}
then results (\ref{eq:Vitesse-x-algebraic}-\ref{eq:Vitesse-y-algebraic})
are valid, the main contribution being related to the singularities
of the Green function.

If the velocity field $U\left(y\right)$ has $M$ stationary points
$y_{m}$ ($U'\left(y_{m}\right)=0$), then the contributions of the
stationary points, generically of order $1/\sqrt{t}$, usually dominate
the contributions of the singularity of the Green function $\mathbf{G}_{k}$,
in integrals like (\ref{eq:velocity_ocillating_integral}). If $\omega_{\infty}\left(y_{m}\right)\neq0$,
the classical stationary phase approximation (see appendix A) would
lead to \[
\mathbf{v}(y,t)\underset{t\rightarrow\infty}{\sim}\sum_{m=1..M}\mathbf{G}_{k}(y,y_{m})\omega_{\infty}(y_{m})\sqrt{\frac{2\pi}{\left|kU''\left(y_{m}\right)\right|}}\exp\left(\frac{i\epsilon_{m}\pi}{4}\right)\frac{\exp(-ikU\left(y_{m}\right)t)}{\sqrt{t}},\]
 where $\epsilon_{m}$ is the sign of $-kU''\left(y_{m}\right)$.

However, a remarkable fact is that for any stationary point $y_{0}$
(such that $U'\left(y_{0}\right)=0$), due to the vorticity depletion
mechanism discussed in section \ref{sub:asymptotic-vorticity-field}
and proved in section \ref{sub:Limit-Stationary-points}, $\omega_{\infty}\left(y_{0}\right)=0$.
Then the leading order contribution from the stationary phase approximation
vanishes. The analysis could proceed in order to determine the next
leading order term in the expansion, from (\ref{eq:velocity_ocillating_integral}),
expected to be of order $1/t^{3/2}$. However, such a detailed analysis
is useless, because the convergence of $\omega(.,t)$ towards the
asymptotic vorticity profile $\omega_{\infty}$ is too slow, in the
vicinity of a stationary streamline. Actually the error due to the
slow convergence towards the Lundgren's profile gives contributions
which are also of order $1/t^{3/2}$. This will be illustrated using
direct numerical simulations in section \ref{sec:Kolmogorov-flow}
(see figure \ref{fig:w1_abs_prof_Lx1_Ly1.1} page \pageref{fig:w1_abs_prof_Lx1_Ly1.1},
figure \ref{fig:w1_abs_prof_Lx1_Ly1.1_i1} page \pageref{fig:w1_abs_prof_Lx1_Ly1.1_i1}
and the related text).

\paragraph{Laplace tools}

In order to give a precise argument for the results (\ref{eq:Vitesse-x-algebraic}-\ref{eq:Vitesse-y-algebraic}),
we use Laplace tools. We first note that $v_{y}=ik\psi$, and study
the asymptotics for the stream function $\psi$. Starting from the inverse
Laplace transform of $\psi$, we have \begin{equation}
\psi\left(y,t\right)=\frac{k}{2\pi}\int_{-\infty}^{+\infty}dc\,\phi_{+}\left(y,c\right)\exp\left(-ikct\right),\label{eq:Inverse_Laplace_psi}\end{equation}
 where $\phi_{+}\left(y,.\right)$ is the limit of $\phi\left(y,c+i\epsilon\right)$
for $\epsilon\rightarrow0$. (\ref{eq:Inverse_Laplace_psi}) is an
oscillating integral. We use that for any $y$ for which $U'\left(y\right)\neq0$,
$\phi_{+}\left(y,c\right)$ is twice differentiable except at $c=U\left(y\right)$
where it has a logarithmic singularity $\Delta\phi_{c}\left(c-U\left(y\right)\right)\log\left(c-U\left(y\right)\right)$
(see equation (\ref{eq:singularite_c})). Then the large time asymptotics
of $\psi$ is due to this singularity. In order to evaluate it, we
part integrate twice (\ref{eq:Inverse_Laplace_psi}) and evaluate
the contribution of the singularity with the residue theorem. Then
the leading order contribution is obtained as, \begin{equation}
\psi\left(y,t\right)\underset{t\rightarrow\infty}{\sim}\frac{\omega_{\infty}\left(y\right)}{\left(ikU'\left(y\right)\right)^{2}}\frac{\exp\left(-ikU(y)t\right)}{t^{2}},\label{eq:asymptotic_psi}\end{equation}
 where we have used (\ref{eq:Delta_Phi_c}) and (\ref{eq:omega_asymptotic_phi})
in order to express $\Delta\phi_{c}$.

We note that (\ref{eq:Vitesse-y-algebraic}) follows immediately from
(\ref{eq:asymptotic_psi}) and from the relation $v_{y}=ik\psi$. The asymptotic result
(\ref{eq:Vitesse-x-algebraic}) for the transverse velocity $v_{x}$
can be derived by following similar arguments as the one just described
for $\psi$.\\

The above argument uses the explicit prediction (\ref{eq:singularite_c})
for the singularity of $\phi_{+}\left(y,c\right)$ as a function of
$c$. The expressions (\ref{eq:singularite_c}) and (\ref{eq:Delta_Phi_c})
are valid only when $y$ is not a stationary streamline ($U'\left(y\right)\neq0$).
For stationary streamlines $y_{m}$ we have no theoretical predictions.
Direct numerical computation in section \ref{sub:numerical-computation}
will lead us to conjecture that for such special points $\psi\left(y_{m},t\right)\underset{t\rightarrow\infty}{\sim}C_{1}\exp\left(-ikU(y_{m})t\right)/t^{3/2},$
$v_{x}\left(y_{m},t\right)\underset{t\rightarrow\infty}{\sim}C_{2}\exp\left(-ikU(y_{m})t\right)/t^{3/2}$
and $v_{y}\left(y_{m},t\right)\underset{t\rightarrow\infty}{\sim}ikC_{1}\exp\left(-ikU(y_{m})t\right)/t^{3/2}$.
We note that this exponent $3/2$ is not related to a contribution
from the stationary phase approximation.\\

\textbf{We thus conclude that the results (\ref{eq:Vitesse-x-algebraic})
and (\ref{eq:Vitesse-y-algebraic}) are valid for monotonic profiles,
but also for base flow with stationary streamlines. This is in marked
contrast to what was thought in many previous publications based on
the asymptotic expansions and the stationary phase arguments.} 
This is mainly due to the vorticity depletion mechanism at the stationary
streamlines discussed in the previous section. We also stress that,
using Laplace tools, the asymptotic profile $v_{\infty}\left(y\right)$
can be numerically computed easily, without relying on direct numerical
computations of the Euler equations.

We have theoretical predictions for the power law in the asymptotic
behavior of the perturbation velocity, for all points of the domain except along the stationary
streamlines.

\section{Asymptotic stability of parallel flows for the Euler equations\label{sec:Asymptotic-stability}}

In the previous section, we have obtained results for the asymptotic
behavior of the linearized Euler equations, with initial conditions
close to some parallel flows $\mathbf{v}_{0}\left(\mathbf{r}\right)=U\left(y\right)\mathbf{e}_{x}$.
We now address the evolution of the same initial conditions by the
nonlinear Euler equation (\ref{eq:Euler}). The aim of this section
is to explain why the linearized dynamics will be a good approximation
for the dynamics for any time $t$, and to explain why the flow velocity
is asymptotically stable (in kinetic energy norm), for small initial
perturbation of the vorticity (in the enstrophy norm).\

We consider the initial vorticity $\Omega\left(x,y,0\right)=-U'\left(y\right)+\epsilon\omega\left(x,y,0\right)$,
where $\epsilon$ is small. We suppose, without loss of generality
that $\int dx\,\omega=0$. The perturbation $\omega$ can be decomposed
in Fourier modes along the $x$ direction \[
\omega\left(x,y,t\right)=\sum_{k}\omega_{k}\left(y,t\right)\exp\left(ikx\right).\]
 From the Euler equations (\ref{eq:Euler}), the equation for the
evolution of $\omega_{k}$ reads \begin{equation}
\frac{\partial\omega_{k}}{\partial t}+ikU\left(y\right)\omega_{k}-ik\psi_{k}U''\left(y\right)=-\epsilon NL\,\,\,\mbox{with}\,\,\, NL=\sum_{l}\left\{ -ik\frac{\partial\psi_{l}}{\partial y}\left(y,t\right)\omega_{k-l}\left(y,t\right)+\frac{\partial}{\partial y}\left[il\psi_{l}\left(y,t\right)\omega_{k-l}\left(y,t\right)\right]\right\} .\label{eq:Euler_k}\end{equation}
 The left hand side is the linearized Euler equation, whereas the
right hand side are the nonlinear corrections. We want to prove that,
for sufficiently small $\epsilon$, neglecting the nonlinear terms
is self-consistent.

For this we have to prove that the nonlinear terms remain uniformly
negligible for large time. We then use the asymptotic results for
the linearized equation derived in section (\ref{sec:theory}). We
thus have, for any $k$\begin{equation}
\psi_{k,L}\left(y,t\right)\underset{t\rightarrow\infty}{\sim}\frac{\omega_{k,L,\infty}\left(y\right)}{\left(ikU'\left(y\right)\right)^{2}}\frac{\exp\left(-ikU(y)t\right)}{t^{2}}\,\,\,\mbox{and}\,\,\,\omega_{k,L}\left(y,t\right)\underset{t\rightarrow\infty}{\sim}\omega_{k,L,\infty}\left(y\right)\exp\left(-ikU(y)t\right),\label{eq:Asymptotics_k}\end{equation}
 where the subscript $L$ refers to the evolution according to the
linearized dynamics. We call a quasilinear approximation to the right
hand side of equation (\ref{eq:Euler_k}), the approximation where
$\psi_{k}$ and $\omega_{k}$ would be evaluated according to their
linearized evolution close to the base flow $U\left(y\right)$. From
(\ref{eq:Asymptotics_k}), one would expect at first sight that this
quasilinear approximation of the nonlinear term $NL_{QL}$, would
give contributions of order $O\left(1/t\right)$. The detailed computation,
easily performed from (\ref{eq:Asymptotics_k}), actually shows that
the contributions of order $O\left(1/t\right)$ identically vanish
for large times. Then \[
\epsilon NL_{k,QL}\underset{t\rightarrow\infty}{=}O\left(\frac{\epsilon}{t^{2}}\right)\]
 This is an important remark, as it proves that within a quasilinear
approximation, the contribution of the nonlinear terms $NL_{QL}$
remains uniformly bounded, and more importantly it is integrable with
respect to time.

Then it is natural to conjecture that the contribution of the nonlinear
terms remains always negligible. More precisely, we naturally conjecture,
that within the fully nonlinear equation, for sufficiently small $\epsilon$:\[
\psi_{k}\left(y,t\right)\underset{t\rightarrow\infty}{\sim}\frac{\omega_{k,\infty}\left(y\right)}{\left(ikU'\left(y\right)\right)^{2}}\frac{\exp\left(-ikU(y)t\right)}{t^{2}}\,\,\,\mbox{and}\,\,\,\omega_{k}\left(y,t\right)\underset{t\rightarrow\infty}{\sim}\omega_{k,\infty}\left(y\right)\exp\left(-ikU(y)t\right)\]
 with \[
\omega_{k,\infty}\left(y\right)=\omega_{k,L,\infty}\left(y\right)+O\left(\epsilon\right)\]

A similar reasoning in order to evaluate the nonlinear evolution for
the profile $U\left(y\right)$ would lead to the conclusion that for
large times \[
\Omega_{0}\left(y,t\right)\underset{t\rightarrow\infty}{\sim}-U_{\infty}'\left(y\right)\,\,\,\mbox{with}\,\,\, U'_{\infty}\left(y\right)=U\left(y\right)+\delta U\left(y\right),\]
 where $\delta U=O\left(\epsilon^{2}\right)$.

This means that the parallel flow will quickly stabilize again towards
another parallel flow which is close to the initial one. \\

A natural question would be to compute the modified profile. The preceding
analysis leads to the quasi-linear expression \begin{equation}
\delta U\left(y\right)=-\epsilon^{2}\int_{0}^{\infty}dt\, NL_{0,QL}\left(t\right)+o\left(\epsilon^{2}\right).\label{eq:Quasilinear_Profile}\end{equation}
 This expression involves integrals over times of the linearized Euler
equation. It is not amenable to a simple expression, but could be easily be computed numerically from Laplace tools.

This result has to be contrasted with the results usually obtained using a quasi-linear approach, for instance in the kinetic theory of particle dynamics (point vortex models, plasma physics, astrophysics). Usually the integrals occurring in (\ref{eq:Quasilinear_Profile})
diverge. Then one invokes a time scale separation, and the divergence of the integral is regularized using a multiple scale analysis. Here,
by contrast, the integral converges. This means that there is a single time scale over which all quantities reach their asymptotic value
(a typical time scale here is $1/s$ where $s$ is the typical shear). The nonlinear evolution is thus very brief and leads to very small
changes in the initial profile $U$.

A theory for the relaxation towards equilibrium for the 2D Euler equations has been proposed based on a quasi-linear theory  coupled to some Markovianization hypothesis, by analogy with the kinetic theory of point vortices \cite{Chavanis_PhRvL_2000_QuasilinearEuler}. Such an approach based on analogies are natural guess that rely on theoretical hypothesis (quasi-linear hypothesis and Markovian hypothesis), that would benefit from either theoretical justifications or numerical verifications.  The results of this paper for the linearized dynamics and the discussion above, show that the approximation of the non-linear dynamics by the linearized dynamics remains uniformly self consistent, and a quasi-linear approach self-consistent. However as discussed above, the relaxation is then extremely rapid and all quantities relax rapidly. Any further assumptions like Markovianization seems then irrelevant. Simpler approach, like the one of \cite{Lundgren_1982_PhFl}, similar to the discussion of this section seems thus more relevant.

We thus conclude that a direct asymptotic expansion based on a quasi-linear approach, similar to the one in \cite{Lundgren_1982_PhFl} or may be with more subtle treatments of the lower order contributions, is most probably the relevant approach for a theory of the relaxation of the 2D-Euler equations. The exact results on the linearized dynamics of the previous section and the discussion above put such an asymptotic expansion on a more rigorous basis, by proving that the leading order contribution remains self-consistently bounded for all times, and explaining why an asymptotic expansion possibly converges. In the Lundgren approach, the asymptotic behavior is described by an asymptotic expansion for large times, whose leading order term is not determined. It actually depends on the initial condition and can not be predicted only with a large time  asymptotic expansion. The further interest of the Laplace method developed in this paper is to give precise predictions for the asymptotic profile that can be easily computed numerically from the Laplace tool. The numerical computations of next section will confirm the statements on this last paragraph, by showing excellent agreements between direct numerical simulation of the Euler equations on one hand and the prediction of such a simple quasi-linear approach, based on Laplace tools computations. \\

We also conclude that, for any profile $U$ verifying the hypothesis of
this work (no unstable and no neutral modes for the linearized dynamics)
with any perturbation corresponding to a small vorticity, the assumption
that the velocity converges for large times towards a new parallel
velocity profile close to the initial profile $U$ is a
self-consistent hypothesis. We see in section \ref{sec:Kolmogorov-flow}
that this is confirmed by numerical computations.

From this discussion, we thus conclude that it is natural to conjecture
that for any profile $U$ verifying the hypothesis of this work (no unstable
and no neutral modes for the linearized dynamics), for any perturbation
corresponding to a small vorticity, the velocity converges
for large times towards a new paralleled velocity profile close
to the initial profile $U$. A possible theorem expressing this more
precisely would be similar to the one recently obtained by Mouhot
and Villani \cite{Mouhot_Villani:2009}, for the Landau damping in
the very close setup of the Vlasov equation. It has however to be noted that
no proof of such a theorem for the Euler equations is known yet,
even in the simplest case of a profile $U$ without stationary points.

It is thus very natural to conjecture that the ensemble of shear flows
without unstable or neutral modes is asymptotically stable %
\footnote{We refer here to the notion of
asymptotic stability of an ensemble of steady states of an infinite dimensional Hamiltonian equations, see for
example the work \cite{Pego_Weinstein_1994CMaPh.164..305P} where any
stable soliton of the KDV equations, slightly perturbed, is proved to converge for large
times towards another slightly different soliton. Asymptotic stability
of ensemble of steady states has also been proved for other solutions of infinite dimensional Hamiltonian
systems.} in the sense given previously (initial perturbation controlled by
a vorticity norm, for instance the enstrophy and large time perturbation
controlled in kinetic energy norm) %
\footnote{A classical argument, presented in a rigorous framework by Caglioti
and Maffei \cite{Caglioti_Maffei_1998} in the context of the Vlasov
equation, implies that steady states of the Vlasov equation for which
Landau damping would occur, would be unstable in a weak norm.\emph{
}At the core of the argument lies the time reversal symmetry of the
equations. These arguments would be easily generalized to the Euler equations.
This may seem in contradiction with the notion of asymptotic stability discussed here.
However the notion of stability discussed by Caglioti and Maffei involves
weak topology for both the initial conditions and final state. There
is no contradiction with our definition of asymptotic stability, as
we control here the initial perturbation in a vorticity norm and control
the convergence in a velocity norm.%
}.

\section{The Kolmogorov flow\label{sec:Kolmogorov-flow}}

In this section, we consider the particular case of the 2D Euler equations
in a doubly periodic domain $\mathcal{D}=[0,2\pi/\delta)\left[0,2\pi\right)$,
where $\delta>1$ is the aspect ratio ; with the Kolmogorov base flow
$U\left(y\right)=\cos\left(y\right)$.

\subsection{Stability\label{sec:Stability}}


In this section we study the stability of the Kolmogorov flow for 2D Euler equation dynamics.
We note that, in the linearized 2D Navier-Stokes equations,
Mishalkin and Sinai \cite{Mishalkin_Sinai_KolmogorovFlow} have found that
the Kolmogorov flow is stable for $\delta>1$ and unstable for $\delta<1$.
Here we prove the stability for $\delta>1$
for the dynamics of both the Euler and linearized Euler equations.
We also show numerically that unstable modes exist for $\delta<1$,
while no stable mode exists for $\delta>1$ for the linearized 2D Euler equations.

\subsubsection{Lyapounov stability}

We consider initial vorticity conditions close to the base flow vorticity
$\omega_{0}\left(y\right)=\sin\left(y\right)$. We let this initial
condition evolve according to the nonlinear Euler equations (\ref{eq:Euler}).
If the perturbation to the initial flow remains small for all times,
the flow is said to be Lyapounov stable. We first prove that the base
flow $\omega_{0}\left(y\right)$ is Lyapounov stable as soon as $\delta>1$.

The classical Energy-Casimir method proposed by Arnold \cite{Arnold_1966}
cannot be applied directly here. Indeed, the Kolmogorov flow does
not verify the hypothesis for any of the two Arnold theorems \cite{Arnold_1966}.
However, we can still prove the stability in this case, by direct
analysis. Let us define the energy-Casimir functional $F$ as \[
F\left[\Omega\right]=\frac{1}{2}\int_{\mathcal{D}}\left(\Omega^{2}-\mathbf{V}^{2}\right).\]
 $F$ being half the enstrophy minus the kinetic energy, it is a conserved quantity for the Euler equations.

We first prove that the base flow $\omega_{0}\left(y\right)=\sin\left(y\right)$
($\mathbf{v}{}_{0}=\cos\left(y\right)\mathbf{e}_{x}$) is a minimum
of $F$. We consider the perturbation vorticity $\omega=\Omega-\omega_{0}$,
and decompose it into Fourier modes $\omega=\sum_{k>1}\omega_{k}e_{k}$
with $\Delta e_{k}=-\lambda_{k}e_{k}$, where the $\lambda_{k}>0$
are arranged in increasing order, and where $e_{k}$ are orthonormal
($\int_{\mathcal{D}}e_{k}e_{k'}=\delta_{kk'}$). On the doubly periodic
domain $\mathcal{D}=[0,2\pi/\delta)\left[0,2\pi\right)$, if $\delta>1$,
we have $\lambda_{1}=\lambda_{2}=1$, corresponding for instance to
the modes $\cos(y)$ and $\sin\left(y\right)$. Then for any $k\geq3$,
$\lambda_{k}>1$.

We obtain \begin{equation}
F=\frac{1}{2}\sum_{k}\frac{\lambda_{k}-1}{\lambda_{k}}\omega_{k}^{2}.\label{eq:Norme_Stability}\end{equation}
Since $\lambda_{k}\geq1$, $F\geq0$. Moreover $F\left[\omega_{0}\right]=0$.
We thus conclude that $\omega_{0}$ is a global minimum for $F$.

We note that this minimum is degenerate, as all vorticity fields
$\omega=\alpha\cos\left(y\right)+\beta\sin\left(y\right)$ are also
minima.

Since $F$ is a conserved quantity, we conclude that \begin{equation}
\frac{1}{2}\sum_{k\geq3}\frac{\lambda_{k}-1}{\lambda_{k}}\omega_{k}^{2}\left(t\right)=\varepsilon_{F},\label{eq:epsilonF}\end{equation}
 where $\varepsilon_{F}=F\left(0\right)$ is the small value of $F$
for the initial perturbation. Then if they are initially small, all
$\omega_{k}$ for $k\geq3$ remain small for large times, the amplitude
being measured according to the norm (\ref{eq:Norme_Stability}).

Expression (\ref{eq:epsilonF}) does not control the first Fourier
modes $\Omega_{1}$ and $\Omega_{2}$. For this, we use the fact that
the enstrophy \[
\Gamma_{2}=\int_{\mathcal{D}}\Omega^{2}\]
 is conserved. We suppose that its initial value is $\Gamma_{2,0}+\epsilon_{\Gamma}$
where $\Gamma_{2,0}$ is the base flow enstrophy and $\epsilon_{\Gamma}$
is the perturbation enstrophy. Using the enstrophy conservation we
have \[
\Omega_{1}^{2}(t)+\Omega_{2}^{2}(t)=\Gamma_{2,0}+\epsilon_{\Gamma}-\sum_{k}\omega_{k}^{2}.\]
 Then, using that $\sum_{k\geq3}\omega_{k}^{2}\leq\frac{2\lambda_{3}}{\lambda_{3}-1}\epsilon_{F}$
(derived from (\ref{eq:epsilonF}), using $\lambda_{k}\geq\lambda_{3}$
for $k\geq3$), we have \[
\left|\Omega_{1}^{2}(t)+\Omega_{2}^{2}(t)-\Gamma_{2,0}\right|\leq\max\left\{ \epsilon_{\Gamma},\frac{2\lambda_{3}}{\lambda_{3}-1}\epsilon_{F}\right\} .\]
 This means that the flow associated to the two first mode is $a(t)\sin\left(y+\phi(t)\right)$
where $\phi$ may be arbitrary but where the amplitude $a$ is controlled
up to an error of order $\max\left\{ \epsilon_{\Gamma},\epsilon_{F}\right\} $.

We have thus proved that any initial condition close to the initial
profile $\omega=\sin\left(y\right)$ remains close to the family of
profiles $\sin\left(y+\phi\right)$. Then the flow is Lyapounov stable
in this sense.

\subsubsection{Linear and spectral stability \label{sub:Linear-stability}}

Next, we let the initial conditions close to the base flow $\omega_{0}(y)=\sin{y}$
evolve according to the linearized 2D Euler equations (\ref{eq:Linearized_Euler}).
If the perturbation to the initial flow remains small for this dynamics,
the flow is said to be linearly stable.

We decompose the perturbation vorticity in Fourier series for the
$x$ variable only. For parallel flows, due to the translational invariance,
such Fourier modes are independent. The modes with no dependence on
$x$ are easily shown to be neutral. Then the only issue is about
the stability of other modes. In order to prove this, we simply note
that the perturbed Energy-Casimir functional (\ref{eq:Norme_Stability})
is conserved not only by the nonlinear Euler dynamics but also by
the linearized one. Then, because the Energy-Casimir functional is
positive, this proves that any $x$-dependent perturbation remains
small if it is initially small. The flow is thus linearly stable as
soon as $\delta>1$.

If the linear equation has no exponentially growing modes, it is called
spectrally stable. Linear stability implies spectral stability (the
converse may be wrong). Then because it is linearly stable, we can
conclude that no unstable modes exist to the linearized 2D Euler equation
as soon as $\delta>1$.

\subsubsection{Neutral modes}

We look for the modes of equation (\ref{eq:Linearized_Euler}) such
that the stream function is of the form $\psi=\phi\left(y\right)\exp\left(ik\left(x-ct\right)\right)$.
Then $\phi$ satisfies the classical Rayleigh equation (\ref{eq:Rayleigh}).

As discussed in section \ref{sub:Linear-stability}, no unstable eigenvalue
exists for $k^{2}>1$. Thus only for real values of $c$ can the Rayleigh
equation have solutions for $k^{2}>1$. In the following, using numerical
simulations, we show that no neutral modes exist, except for the marginal
case $k=1$.\\

When $c$ is in the range of $U$: $-1=\min_{y}\left\{ U\left(y\right)\right\} <c<1=\max_{y}\left\{ U\left(y\right)\right\} $,
$U-c$ vanishes at the two critical layers defined by \textbf{$U\left(y_{l_{1,2}}\right)=\cos\left(y_{l_{1,2}}\right)=c$}.
The Rayleigh equation then has logarithmic singularities. As discussed
in section \ref{sub:Initial_value_problems}, when initial value problems
are considered, the relevant solutions to the Rayleigh equation are
the ones that are obtained with $c'=c+i\epsilon$, $c$ real, and
by considering the limit $\epsilon\rightarrow0^{+}$. We study the
existence of modes in that limit.

For this, we numerically compute the dispersion relation $D_{+}\left(c,k\right)$
of the Rayleigh equation, as defined in section \ref{sub:Single-Critical-Layer}.
Neutral modes correspond to zeros of $D_{+}$. We use the same numerical
tools as the one described in the end of section \ref{sub:Single-Critical-Layer}: we use the Matlab function \emph{ode45,} and fix the relative
error parameter and the absolute error parameter of this function to $10^{-13}$,
then obtain solutions for which errors in the Wronskian $W$ are typically
smaller than $10^{-6}$ for $\epsilon=10^{-4}$. We approximate $D_{+}\left(c,k\right)$
by the numerically computing $D\left(c+i\epsilon,k\right)$ with $\epsilon=10^{-4}$.

\begin{figure}
\begin{centering}
\includegraphics[width=0.75\columnwidth]{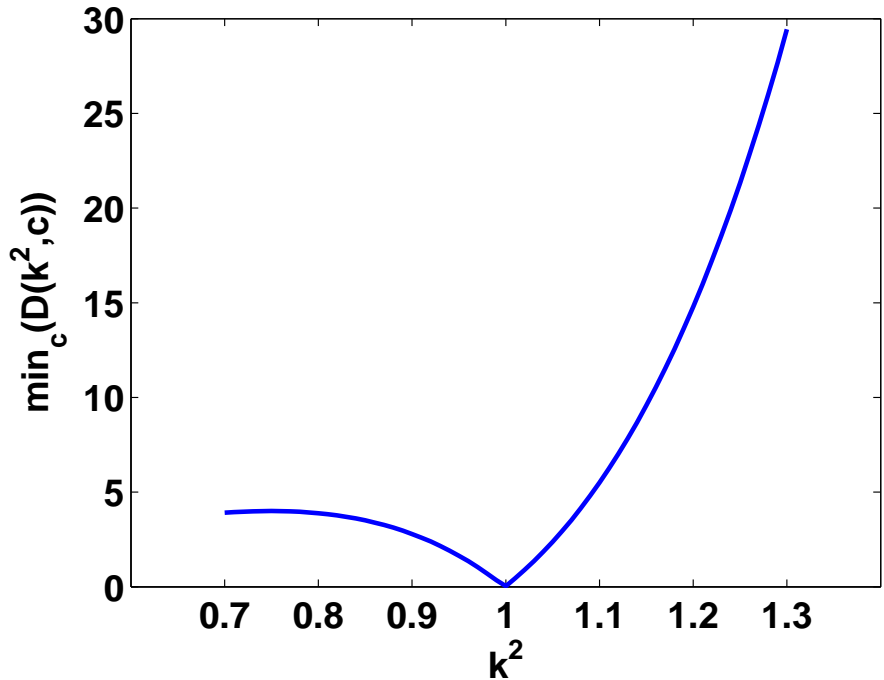}
\par\end{centering}

\caption{\label{fig:dispersion} Minimum values for the dispersions relation
$\min_{c}D^{+}\left(c,k\right)$ as a function of $k^{2}$. This plot
shows that neutral modes exist only for $k^{2}=1$. }

\end{figure}

Figure \ref{fig:dispersion} shows $D_{m}\left(k\right)=\min_{c}D^{+}\left(c,k\right)$
(we note that $D^{+}$ is unchanged when the sign of $k$ is changed).
For a given value of $k$, some neutral mode exist if and only if
$D_{m}\left(k\right)$ vanishes. We conclude from this plot that neutral
modes exist only for the value $k^{2}=1$ (we have tested values of
$k^{2}$ up to $k^{2}=10$).

\begin{figure}
\begin{centering}
\includegraphics[width=0.75\columnwidth,keepaspectratio]{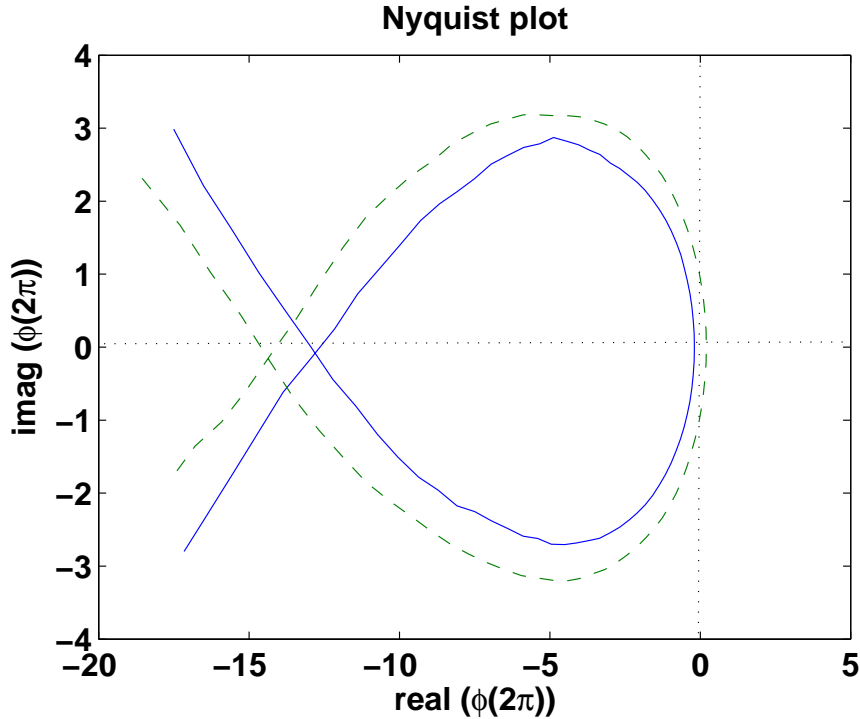}
\par\end{centering}

\caption{\label{fig:Nyquist} Nyquist plots (complex $D^{+}\left(c\right)$
represented in the complex plane, when $c$ is varied), for values
$k^{2}=0.99$ (green dashed line) and $k^{2}=1.01$ (plain blue line). }

\end{figure}

For $k^{2}=1$, we see numerically that a mode exists for $c=0$ only.
The mode is then found by direct integration of equation
(\ref{eq:Rayleigh}). It is the trivial mode $\phi=C$ where $C$ is any constant number
($\psi=C\exp\left(ikx\right)$ and $\psi=C\exp\left(-ikx\right)$).\\

The representation of the complex curve $D_{+}^{}\left(c\right)$
when $c$ is varied is called a Nyquist plot (see \cite{Nicholson_1991}
or \cite{Balmforth_Morrison_1999} in the context of fluid dynamics).
It is very useful, as the algebraic number of loops of the Nyquist
plot around $0$ counts the number of unstable modes on the complex
half plane $c=c_{R}+i\lambda$ with positive $\lambda$ \cite{Nicholson_1991,Balmforth_Morrison_1999}.

Figure \ref{fig:Nyquist} shows the Nyquist plot of $D_{+}\left(c\right)$,
for $k=0.99$ and $k=1.01$ respectively. One clearly sees the passing
of the curves across the value $D_{+}=0$ when $k$ is changed from
$k=0.99$ to $k=1.01$, corresponding to neutral modes for $k=1$.
Moreover, we conclude that only one neutral mode exists for this value
of $k$, because only one branch of the curve passes through $0$.
For larger value of $c$ (not shown), the upper part of the plot loops
to the right on the upper half plane goes down to the lower half plane
by crossing the real axis for very large values of $\phi$, before
to close on the branch visible in the lower half plane. Counting the
algebraic number of loops around zero, we thus conclude that for $k>1$,
no unstable modes exists, in accordance with the result of section
(\ref{sec:Stability}) ; whereas for $k<1$ only one unstable mode
exists.

From this analysis, we thus conclude that only one neutral mode exists.
This modes corresponds to the destabilization of the flow and the
appearance of an unstable mode, when passing from values $k^{2}\geq1$
to values $k^{2}\leq1$. It is the trivial mode $\psi=C\exp\left(ix\right)$,
found for the phase speed $c=0$.

\subsection{Direct numerical computation of the 2D-Euler equations\label{sub:numerical-computation}}

In this section, we illustrate and complement the above results through
the direct numerical simulation of the 2D (nonlinear) Euler equations,
in the doubly periodic domain $\mathcal{D}=[0,2\pi/\delta)\left[0,2\pi\right)$,
for the Kolmogorov base flow $U(y)=\cos y$.

Since the base flow is homogeneous in the $x$ direction,
the dynamics of the fields would be decoupled into that of the components
of the form $f_k(y,t)\exp\left(ikx\right)$,
if the equations were linearized. Then it is natural to consider fields of this form
also in the (nonlinear) Euler equations, because we are interested in slightly perturbed parallel flows.

Since, in the Euler equations, instabilities are mainly large scale ones,
the flow is more likely affected by large scale perturbations.
Moreover, in geophysics and experiments, forcing and perturbations tend to be dominantly effective
on the largest scales of the flow.
It is thus natural to study initial perturbations of the form
\begin{align}
\omega(x,y,0) & =\epsilon A(y)\cos\delta x
\end{align}
where $k=\delta$ is the smallest wavenumber (largest scale).
We consider the case $\epsilon\ll1$ ;
we use $\epsilon=0.01$ throughout in the following numerical computations.

Similarly, we first examine the dynamical response for the same wave number as the initial perturbation, namely,
\begin{align}
\omega_{\delta}(y,t) & =\int\frac{\d x}{2\pi\delta^{-1}}\;\e^{-\i\delta x}\omega(x,y,t)\\
\bm{v}_{\delta}(y,t) & =\int\frac{\d x}{2\pi\delta^{-1}}\;\e^{-\i\delta x}\bm{v}(x,y,t).
\end{align}
The analysis of nonlinear effect will be performed at the end of this section.

\begin{figure}
\begin{center}
\includegraphics[width=0.9\textwidth]{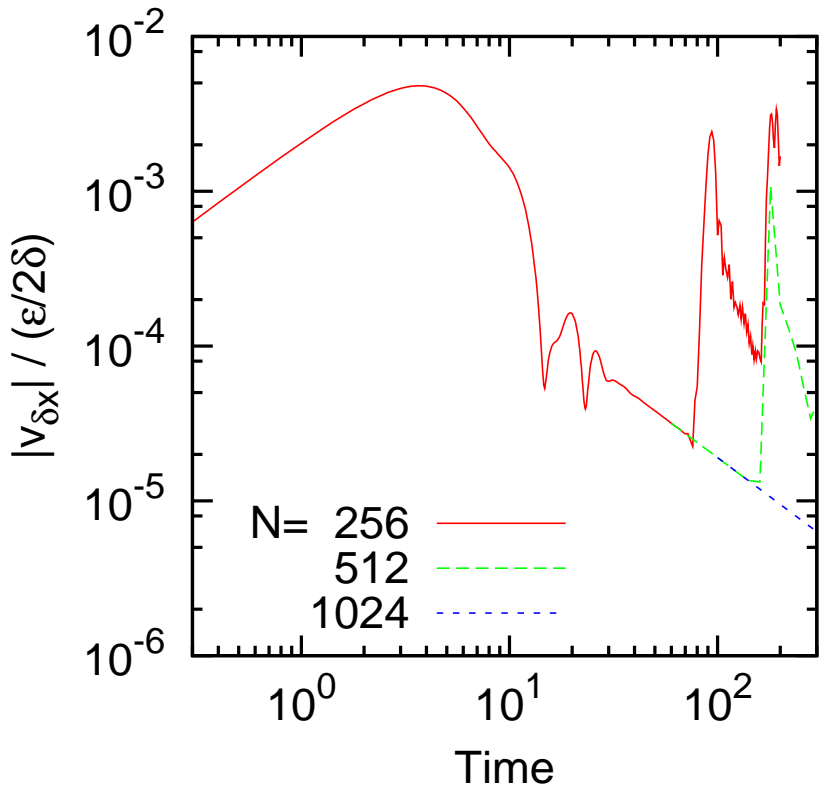}
\end{center}
\caption{
The time series of perturbation velocity components $|v_{\delta,x}(y=0,t)|$,
for the initial perturbation profile $A(y)=1$ and the aspect ratio $\delta=1.1$,
with various resolutions of the system ($N\times N$).
The simulations blow up at $t\approx 70$ for $N=256$
and at $t\approx 150$ for $N=512$.
In the following numerical results,
we have used high resolution enough to confirm
that the asymptotic behavior is robustly observed.
}
\label{fig:vx1_abs_tseri_Lx1_Ly1.1_vs_N.eps}
\end{figure}

\paragraph{Direct numerical simulations.}

In the following, for the direct numerical simulations, we use the classical pseudo-spectral method algorithm \cite{Orszag},
which is the most precise and robust numerical algorithm currently known for the Euler and Navier Stokes equations in doubly periodic domains,
and indeed regarded as the standard method for precise numerical simulations in these cases.

A standard way to compute numerical solutions to the Euler equations is to integrate directly the truncated Euler equations without viscosity
\cite{Frisch_Matsumoto_Bec_2002}. We follow this procedure. The numerical scheme is thus not stable in the long run; on account of the lack of explicit dissipation, the simulation blows up when the small scale structures become of the same size as the grid scale \cite{Frisch_Matsumoto_Bec_2002}.
As an illustration, figure 8 presents the computation of the first mode of the velocity perturbation, computed with three different numerical simulations using $256^2$, $512^2$ and $1024^2$ Fourier components, respectively. This illustrates the blowup after some times, occurring later on for larger resolutions. Moreover this shows that the short time numerical results are stable when the resolution is increased. In the following numerical results, we have always confirmed that the resolution of the system is high enough, by systematically testing the stability of the results by changing the resolution, such that the asymptotic behavior discussed is robustly observed. This procedure for assessing the numerical results for the Euler equations without dissipation is also a standard one (see for instance \cite{Frisch_Matsumoto_Bec_2002}).

An independent assessment of the quality of the numerical results is provided by figure  \ref{fig:w1_abs_fin_num_theor}, discussed later in more details. On this figure, the final vorticity profile is computed by two algorithms: the just described direct numerical simulations, and the predictions from the Laplace transform (equation (34)). The results from these two completely independent algorithms are indistinguishable on figure  \ref{fig:w1_abs_fin_num_theor}. The maximum discrepancy between the two profiles on this figure is of order 0.0001.

\paragraph{Asymptotic vorticity profile for even perturbations.}

First, we consider an initial perturbation where $A(y)$ has the same
parity as the base flow. In particular, we examine $A(y)=1$.

The space-time series of $|\omega_{\delta}(y,t)|$ is shown in \reffig{fig:Vorticity_Depletion},
page \pageref{fig:Vorticity_Depletion}, which we have already seen.
Initially, it rapidly (almost exponentially) relaxes toward the final
profile, $|\omega_{\delta\infty}(y)|$; in particular, it relaxes
to zero at $y=0$ and $\pi$ (stationary streamlines), whereas it
remains constant at $y=\pi/2$ and $3\pi/2$. The rapid relaxation
of the modulus $|\omega_{\delta}(y,t)|$ is in agreement with the
theoretical prediction (\ref{eq:omega_asymptotic}) that the Lundgren
ansatz is asymptotically valid.

\begin{figure}[H]
\begin{center}
\includegraphics[width=0.9\textwidth]{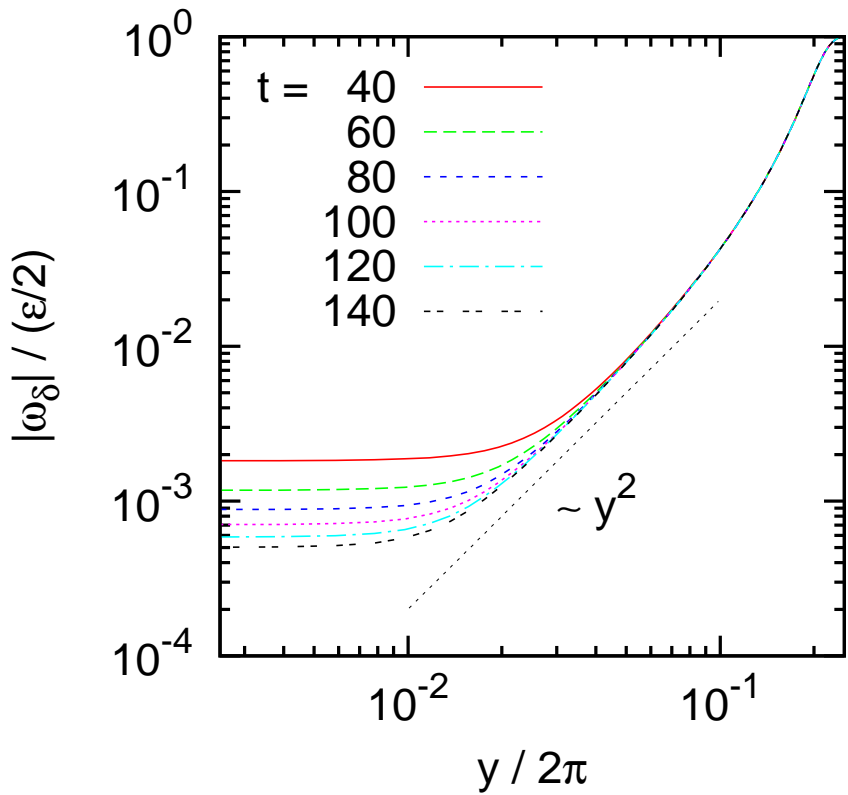}
\end{center}
\caption{ The profiles of perturbation vorticity, $|\omega_{\delta}(y,t)|$,
at several times, for the initial perturbation profile $A(y)=1$ and
aspect ratio $\delta=1.1$.
A flat region is observed near the stationary streamline $y=0$.
As time goes on, this flat region becomes narrower and narrower,
and instead the region with the profile proportional to $y^{2}$
extends towards the stationary streamlines,
leading to a parabolic profile in the large time limit.}
\label{fig:w1_abs_prof_Lx1_Ly1.1}
\end{figure}

\begin{figure}[p]
\begin{center}
\includegraphics[width=0.75\textwidth]{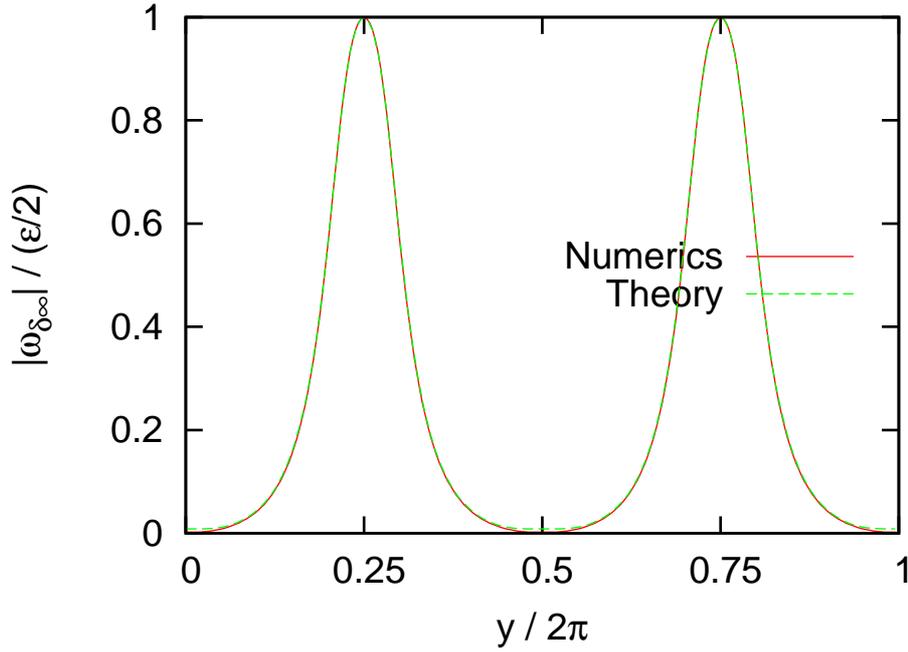}
\end{center}
\caption{The final profile of the modulus of the perturbation vorticity, $|\omega_{\delta\infty}(y)|$,
obtained from the direct numerical simulation  and the theory (equation
(\ref{eq:omega_asymptotic_phi})), for the initial perturbation profile
$A(y)=1$ and the aspect ratio $\delta=1.1$. The two profiles show
a very good agreement. }
\label{fig:w1_abs_fin_num_theor}
\end{figure}

\begin{figure}[p]
\begin{center}
\includegraphics[width=0.75\textwidth]{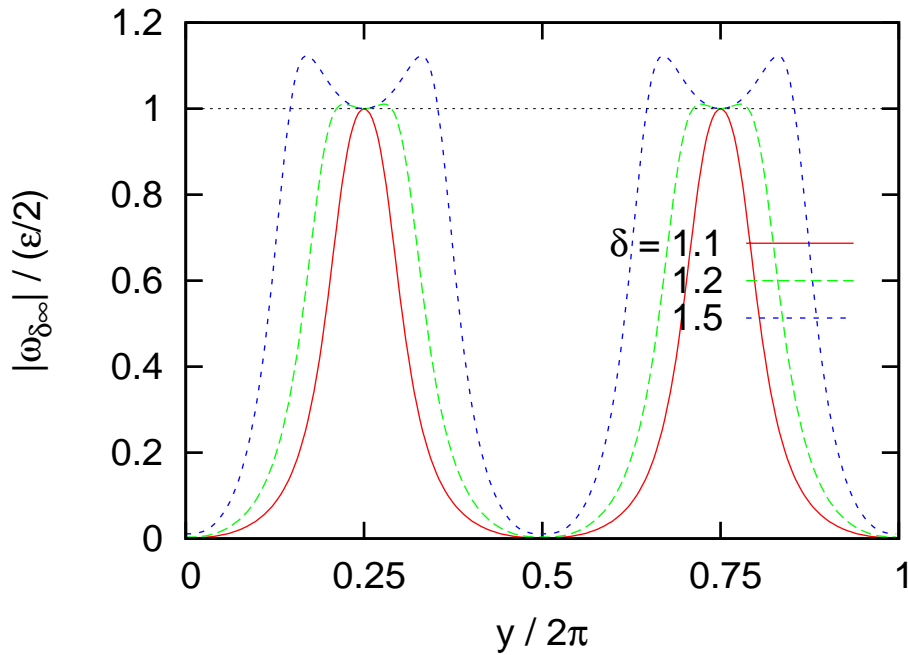}
\end{center}
\caption{ The final profile for the modulus of the perturbation vorticity,
$|\omega_{\delta\infty}(y)|$, for the initial perturbation profile
$A(y)=1$, and aspect ratios $\delta=1.1,1.2$ and $1.5$, computed
from the prediction of the Laplace transform tools (equation (\ref{eq:omega_asymptotic_phi})).
The profile shows a bifurcation from single to double peak shapes,
when $\delta$ is increased. }
\label{fig:w1_abs_fin_vs_aratio}
\end{figure}

\begin{figure}[p]
\begin{center}
(a) \includegraphics[width=0.9\textwidth]{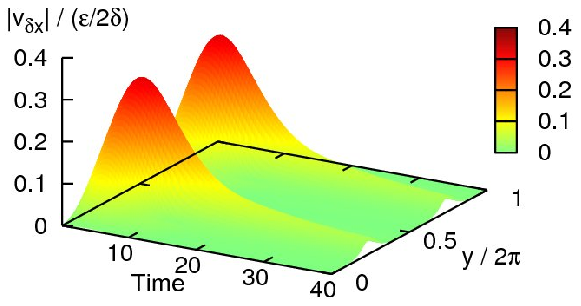}
\\
(b) \includegraphics[width=0.9\textwidth]{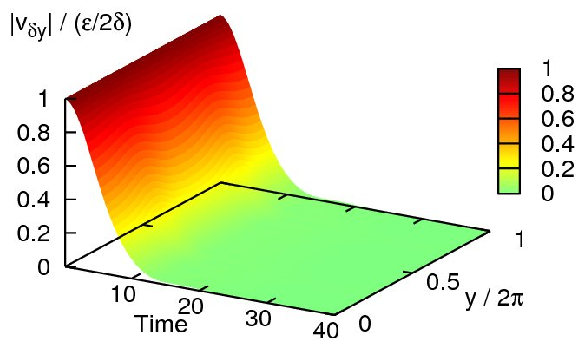}
\end{center}
\caption{ The space-time series of  $|v_{\delta,x}(y,t)|$
(a) and $|v_{\delta,y}(y,t)|$ (b), for the initial perturbation profile
$A(y)=1$ and the aspect ratio $\delta=1.1$. Both the components
relax toward zero, showing the asymptotic stability of the Euler equations. }
\label{fig:v1_stseri}
\end{figure}

\begin{figure}[p]
\begin{center}
(a) \includegraphics[width=0.9\textwidth]{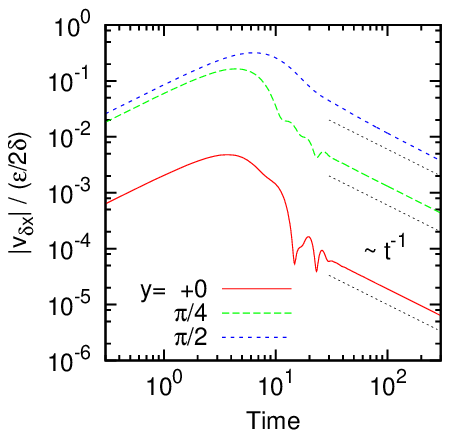}
\\
(b) \includegraphics[width=0.9\textwidth]{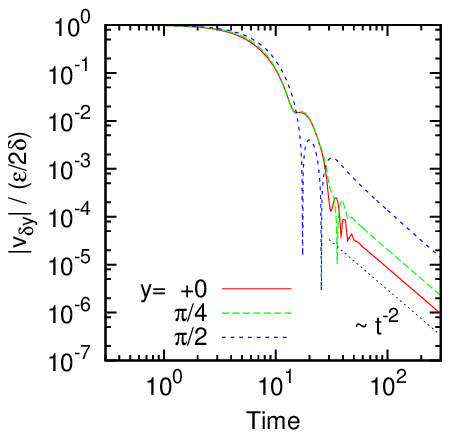}
\end{center}
\caption{The time series of perturbation velocity components $|v_{\delta,x}(y,t)|$
(a) and $|v_{\delta,y}(y,t)|$ (b) at three locations, $y=0$ (vicinity
of the stationary streamline) (red), $y=\pi/4$ (green), and $y=\pi/2$
(blue), for the initial perturbation profile $A(y)=1$ and the aspect
ratio $\delta=1.1$. We observe the asymptotic forms $|v_{\delta,x}(y,t)|\sim t^{-\alpha}$,
with $\alpha=1$, and $|v_{\delta,y}(y,t)|\sim t^{-\beta}$, with
$\beta=2$, in accordance with the theory for the asymptotic behavior
of the velocity (equations (\ref{eq:Vitesse-x-algebraic}) and (\ref{eq:Vitesse-y-algebraic})) }
\label{fig:v1_tseri}
\end{figure}

After the rapid relaxation, $|\omega_{\delta\infty}(y)|$ converges
towards the final profile algebraically. In particular, in the vicinity
of $y=0$ and $\pi$ (stationary streamlines), it relaxes as slowly
as $t^{-1}$, leading locally to a flat profile (see figure \ref{fig:w1_abs_prof_Lx1_Ly1.1}).
However, as time goes on, this flat region becomes narrower and narrower,
and instead the region with the profile proportional to $y^{2}$
extends towards the stationary streamlines.
This indicates that, in the large time limit,
the profile is parabolic in the vicinity of the stationary streamlines.

This also illustrates that the relaxation towards the stationary profile
does not converge in a uniform way; the process is very slow close
to the stationary streamlines whereas it is extremely rapid away from
them.

The width $y_{f}$ of the flat area decreases as $y_{f}\sim1/\sqrt{t}$,
whereas the constant value of the vorticity modulus in the flat area
decreases as $1/t$. When computing the velocity from the vorticity
(equation (\ref{eq:Evolution_velocity})), the overall effect of this
flat area is thus of order $1/t^{3/2}$. Such a contribution is thus
of the same order as what would give the leading order term of the
asymptotic expansion of oscillating integrals, as discussed in the
end of section \ref{sub:asymptotic-velocity}.

In \reffig{fig:w1_abs_fin_num_theor}, we compare the final profiles
obtained from the Laplace tools (equation (\ref{eq:omega_asymptotic_phi}))
and the direct numerical simulations. As shown, the results show a
very good agreement. This agreement support both the quality of the
direct numerical simulations and the results of the computation of
the asymptotic profile from the Laplace transform tools.

The computation of the asymptotic profile from the Laplace method
is extremely rapid and easy, compared with direct numerical simulations.
Using this tool, we study some qualitative properties of the asymptotic
profile. By increasing the aspect ratio $\delta$, we observe a bifurcation
from a single- to a double-peaked asymptotic vorticity profile (figure \ref{fig:w1_abs_fin_vs_aratio}).
The three asymptotic profiles all show the depletion of the vorticity
perturbation at the stationary streamlines.

\paragraph{Asymptotic decay of the velocity perturbation for even perturbations.}

The space-time series of the modulus of the perturbation velocity
components $|v_{\delta,x}(y,t)|$ and $|v_{\delta,y}(y,t)|$ are shown
in \ref{fig:v1_stseri}. The relaxation to zero of the velocity
perturbation illustrates the asymptotic stability of the velocity
for the 2D Euler dynamics.

We investigate the asymptotic behavior of the velocity perturbation
more precisely. \reffig{fig:v1_tseri} shows the time series at several positions.
As shown, their asymptotic forms are $|v_{\delta x}(y,t)|\sim t^{-\alpha}$,
with $\alpha=1$, and $|v_{\delta y}(y,t)|\sim t^{-\beta}$, with
$\beta=2$. This is in agreement with the theoretical predictions
for the asymptotic behavior of the velocity perturbation (see equations
\ref{eq:Vitesse-x-algebraic} and \ref{eq:Vitesse-y-algebraic})).

\paragraph{Odd perturbations.}

Next, we consider initial perturbations where $A(y)$ has a parity
opposite to the base flow one. In particular, we examine $A(y)=\sin{y}$.

\begin{figure}[hb]
\begin{center}
\includegraphics[width=0.9\textwidth]{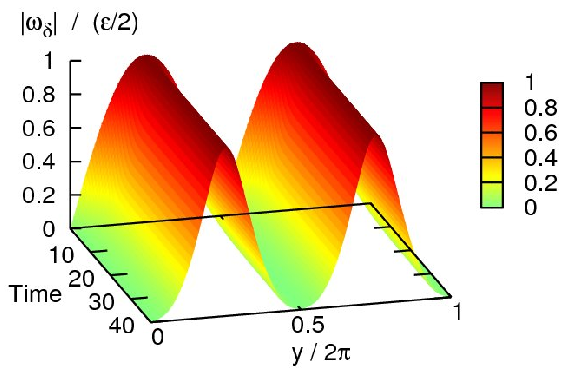}
\end{center}
\caption{ The space-time series of the modulus of the perturbation vorticity,
$|\omega_{\delta}(y,t)|$, with initial perturbation profile $A(y)=\sin{y}$,
and aspect ratio $\delta=1.1$. }
\label{fig:w1_stseri_i1}
\end{figure}

The space-time series of $|\omega_{\delta}(y,t)|$ is shown in \reffig{fig:w1_stseri_i1}.
It shows an initial rapid relaxation toward the final profile, as
expected from the theory (equation (\ref{eq:omega_asymptotic_phi})).
Since the parity of the perturbation is conserved for all time,
the vorticity profile remains odd. Then $|\omega_{\delta}(y,t)|$
is zero for $y=0$ and $y=\pi$ (stationary streamlines), as expected.

\begin{figure}[p]
\begin{center}
\includegraphics[width=0.75\textwidth]{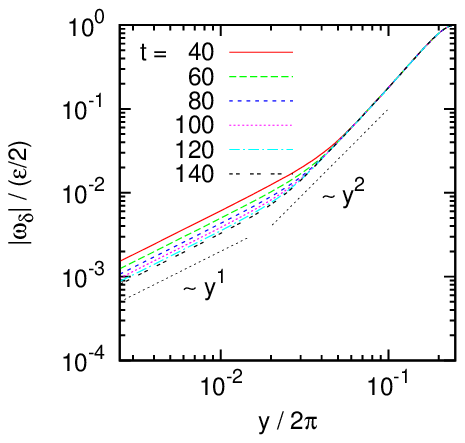}
\end{center}
\caption{ The profiles of the perturbation vorticity modulus, $|\omega_{\delta}(y,t)|$,
at several times, for the initial perturbation profile $A(y)=\sin{y}$
and aspect ratio $\delta=1.1$. As the time goes on, the linear region
becomes narrower and narrower,
and instead the region with the profile proportional to $y^{2}$
extends towards the stationary streamlines,
leading to a parabolic profile in the large time limit.}
\label{fig:w1_abs_prof_Lx1_Ly1.1_i1}
\end{figure}

The rapid relaxation is again followed by an algebraic convergence
to the final profile. In particular, in the vicinity of $y=0$ and
$y=\pi$ (stationary streamlines), it relaxes as slowly as, $t^{-1/2}$,
in this case. The vorticity is always zero at the stationary streamlines,
the profile in the vicinity is linear (see figure \ref{fig:w1_abs_prof_Lx1_Ly1.1_i1}),
not flat as in the case of even perturbations. However, as time
goes on, this linear region becomes narrower and narrower,
and instead the region with the profile proportional to $y^{2}$
extends towards the stationary streamlines.
This indicates that, in the large time limit, the profile
is locally parabolic in the vicinity of the stationary streamlines,
as in the case of even perturbations. The profile being odd, we remark
that such a parabolic profile means that the asymptotic vorticity
profile is not twice differentiable at the stationary streamlines.

\begin{figure}[p]
\begin{center}
\includegraphics[width=0.75\textwidth]{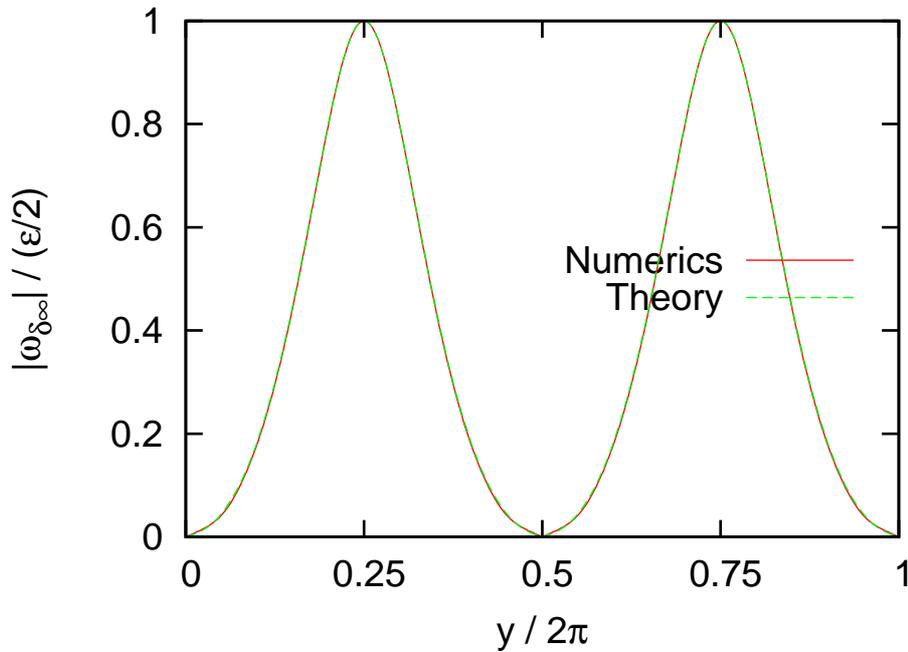}
\end{center}
\caption{The final profile of the perturbation vorticity modulus, $|\omega_{\delta\infty}(y)|$,
obtained from both direct numerical simulation  and theory. The initial
perturbation profile is $A(y)=\sin{y}$ and the aspect ratio is $\delta=1.1$.
The two profiles show excellent agreement. }
\label{fig:w1_abs_fin_num_theor_i1}
\end{figure}

\begin{figure}[p]
\begin{center}
(a) \includegraphics[width=0.9\textwidth]{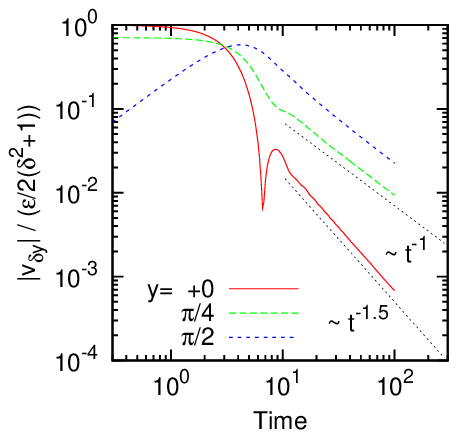}
\\
(b) \includegraphics[width=0.9\textwidth]{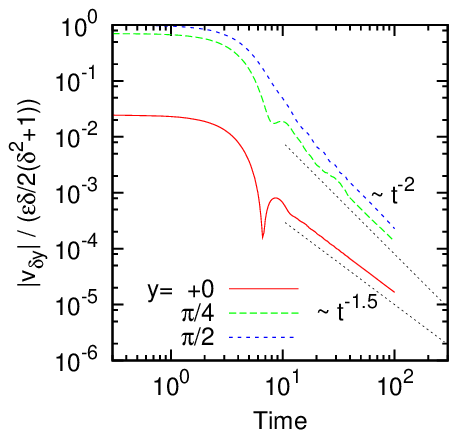}
\end{center}
\caption{ The time series of the perturbation velocity components, $|v_{\delta,x}(y,t)|$
(a) and $|v_{\delta,y}(y,t)|$ (b) at three locations, $y=0$ (vicinity
of the stationary streamline) (red), $y=\pi/4$ (green), and $y=\pi/2$
(blue), for the initial perturbation profile $A(y)=\sin{y}$ and aspect
ratio $\delta=1.1$. We observe the asymptotic forms $|v_{\delta,x}(y,t)|\sim t^{-\alpha}$,
with $\alpha=1$, and $|v_{\delta,y}(y,t)|\sim t^{-\beta}$, with
$\beta=2$, in the almost all the region, in accordance with the theory.
Only in the vicinity of $y=0$ and $\pi$, we observe the exponents
$\alpha=1.5$ and $\beta=1.5$, for which we have no theoretical predictions. }
\label{fig:v1_tseri_i1}
\end{figure}

\begin{figure}[p]
\begin{center}
\includegraphics[width=0.9\textwidth]{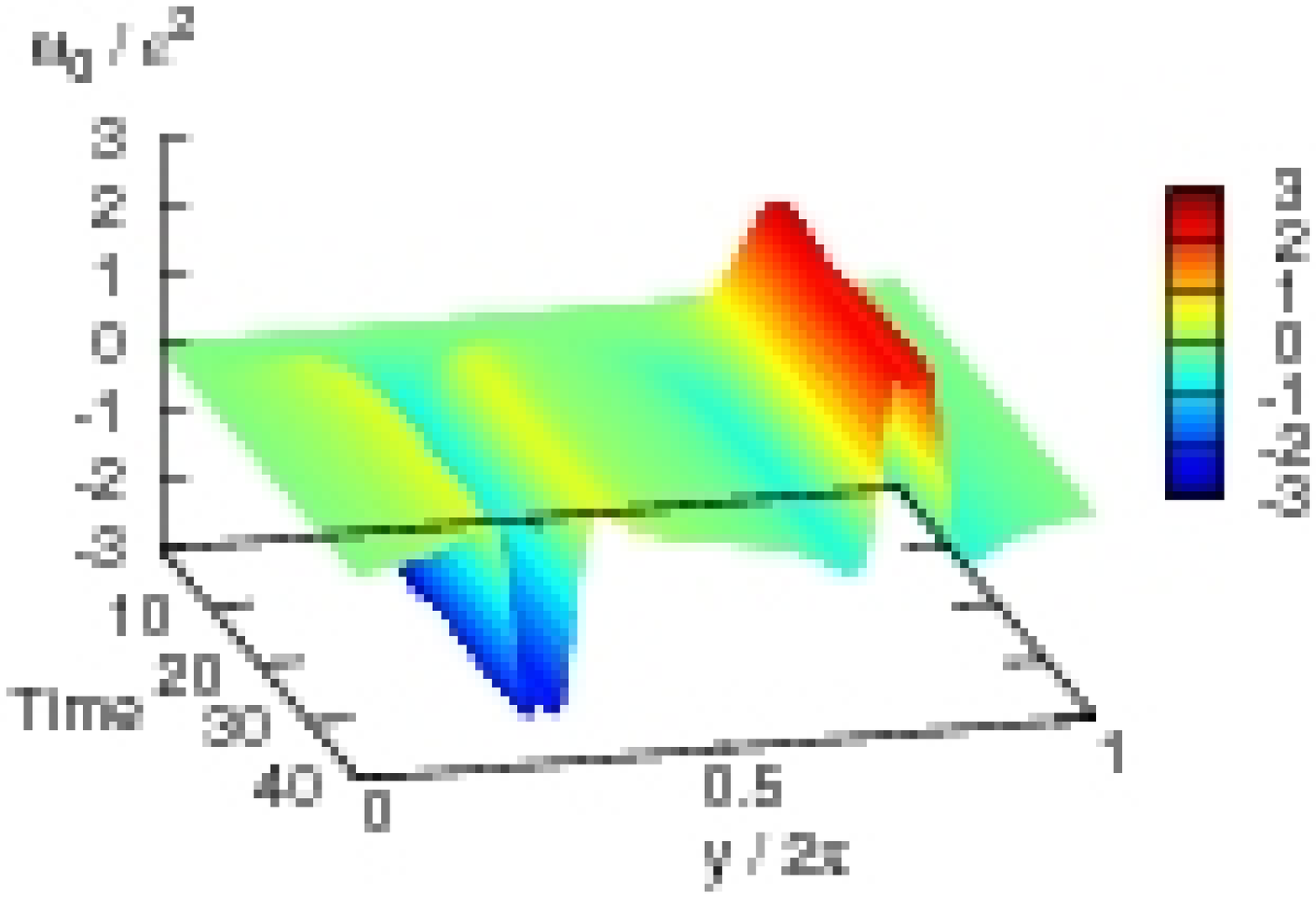}
\end{center}
\caption{ The space-time series of the $x$-averaged perturbation vorticity,
$\omega_{0}(y,t)$. The initial perturbation profile is $A(y)=1$
and the aspect ratio is $\delta=1.1$. }
\label{fig:w0_stseri}
\end{figure}

The final profile obtained from the Laplace transform tools (equation
(\ref{eq:omega_asymptotic_phi})) and the direct numerical simulations
again show excellent agreement (see figure \ref{fig:w1_abs_fin_num_theor_i1}).

The space-time series of $|v_{\delta,x}(y,t)|$ and $|v_{\delta,y}(y,t)|$,
similarly to the case of $A(y)=1$, shows a relaxation toward zero,
illustrating the asymptotic stability of the Euler equations.

We investigate the velocity asymptotic behavior more precisely.
\reffig{fig:v1_tseri_i1}
shows the time series at several positions. As shown, their asymptotic
forms are $|v_{\delta x}(y,t)|\sim t^{-\alpha}$, with $\alpha=1$,
and $|v_{\delta y}(y,t)|\sim t^{-\beta}$, with $\beta=2$, for almost
all values of $y$. Thus we verify that the direct numerical simulation for the
asymptotic behavior of the velocity perturbation is in good agreement
with the theory. Only in the vicinity of $y=0$ and $\pi$ (stationary
streamlines), the exponents are changed to $\alpha=1.5$ and $\beta=1.5$.
We recall that we have no theoretical predictions for the asymptotic velocity on these stationary streamlines. 

In the last paragraphs, we have compared the mode $\omega_{\delta}$
from the direct numerical simulations of the 2D (nonlinear) Euler equations
with the predictions of the 2D linearized Euler equations.
The agreement between both is extremely good.
There is indeed no visible differences, even for large times.
This is in agreement with the theoretical discussion in the section
\ref{sec:Asymptotic-stability}, namely, the difference is
expected to be of the order $\epsilon^{2}$. To summarize, we
conclude that the asymptotic behavior of the (nonlinear) Euler equations
are very well described by the quasi-linear theory
discussed in the previous sections.

\paragraph{Evolution of the base flow profile and asymptotic stability.}

We now consider the evolution of the base flow profile $\Omega_{0}(y,t)$
(the $x$-average vorticity, directly related to the $x$-average
velocity, please see the discussion preceding equation (\ref{eq:Quasilinear_Profile})
page \pageref{eq:Quasilinear_Profile})). We define the difference
with respect to the initial profile by $\omega_{0}(y,t)=\Omega_{0}(y,t)+U'(y)$.
We thus have
\begin{align}
\omega_{0}(y,t)=\int\frac{\d x}{2\pi\delta^{-1}}\omega(x,y,t).
\end{align}

The evolution of the base profile $\omega_{0}(y,t)$ is due
to the nonlinear effects (see Eq. (\ref{eq:Quasilinear_Profile}) page
\pageref{eq:Quasilinear_Profile}). Figure \ref{fig:w0_stseri} shows
this evolution in direct numerical simulations. This illustrates
that the perturbation vorticity converges extremely rapidly (on times
of order $t=15$ which correspond to the linear shear times) toward
a fixed perturbation profile. The asymptotic perturbation profile
is of order $\epsilon^{2}$. All these are in agreement with the theoretical
discussions of section \ref{sec:Asymptotic-stability}.

\section{Discussion\label{sec:Discussion}}

In this paper, we have discussed the asymptotic stability of parallel
flows for the 2D Euler equations. Our results are valid for any flow
that have no modes for the linearized dynamics (neither unstable nor
neutral ones). This situation is a generic one, as the example of
the Kolmogorov flows (section \ref{sec:Kolmogorov-flow}) illustrates.
An adaptation of the present results to the case where the flow has
neutral modes would be easy. Our results are valid for base flow profiles
$U\left(y\right)$ with or without stationary points $y_{0}$
such that $U'\left(y_{0}\right)=0$. We have emphasized the case with
stationary points that has not been studied before.

For the linearized Euler equation, we have proven that Lundgren's
ansatz (\ref{eq:Lungren}) actually describes the asymptotic vorticity
field for large times. The asymptotic vorticity field thus oscillates,
for each streamline, at a multiple of the streamline frequency. The asymptotic vorticity
profile depends both on the initial condition in a non trivial way,
and on the base flow. The asymptotic vorticity is always strongly
affected by the base flow structure, in a non-local way, especially
when stationary streamlines exist. It is thus unlikely that a description
based on the local shear give a good quantitative description, except
may be in a limit or nearly linear shear. We have also shown that
this asymptotic profile can be computed directly from the resolvent
operator of the linearized Euler equation (see equation (\ref{eq:omega_asymptotic_phi})
and figure \ref{fig:w1_abs_fin_vs_aratio}) without performing costly
direct numerical computation of the Euler equations.

For the linearized Euler equations, we have also proved that the asymptotic
velocity field decays algebraically for large times (equation \ref{eq:Vitesse_Asymptotics}),
with exponents $\alpha=1$ and $\beta=2$ for any streamlines that
are not stationary ($U'\left(y_{0}\right)\neq0$). On the stationary
streamlines, we have no theoretical predictions, but we have found
numerically that two cases exist $\alpha=1$ and $\beta=2$ for perturbation
vorticity fields having the same periodicity as the base flow velocity
or $\alpha=3/2$ and $\beta=3/2$ for perturbation vorticity field
having the opposite periodicity with respect to the base flow. Without
stationary streamlines, these results are the same as the classically
expected ones. With stationary streamlines, these results were unexpected as
the effect of the stationary streamlines in oscillating integrals
could have been expected to give $1/\sqrt{t}$ contributions. Such contributions cancel out
because of a self-consistent vorticity depletion at the stationary
streamline. This is a new mechanism of vorticity depletion at the
stationary streamline that we theoretically predict for the linearized 2D Euler equations and prove to be self consistent for the 2D Euler equations,  and numerically confirm for the 2D Euler equations in this paper.

This vorticity depletion mechanism occurs due to the effects of the
transverse component of the velocity perturbation on the background
vorticity gradient. This mechanism is thus absent in cases where the
background vorticity gradient identically vanishes, or for a beta-plane
barotropic flow when the beta effect exactly balance the vorticity
gradient, a case studied in several papers \cite{Brunet_Warn_1990_JAtS,Brunet_Haynes_1995_JAtS}.
We think that this last case is not generic as the vorticity depletion
mechanism exists as soon as the vorticity gradient is not exactly
balanced.

We use the above results to prove that if the perturbation evolves
according to the linearized 2D Euler equations, the nonlinear term remains
uniformly bounded in time, and actually decays algebraically for large
times. Based on these results, we argue that for the Euler equations
(nonlinear), a quasilinear treatment of the nonlinear terms is self
consistent. This strongly suggests that such a quasilinear treatment
of the nonlinear term should be valid. This also suggests that the
full nonlinear equation converges towards Lundgren's type asymptotics
for the perturbation vorticity field and to zero for the asymptotic
velocity field, extremely rapidly.

From these theoretical arguments, we then expect that the velocity
of parallel flows without unstable or neutral linear modes is asymptotically
stable : the velocity converges towards a new parallel flow which
is very close to the initial one, even in the absence of dissipation.
The distance between the initial profile and the asymptotic one is
of order $\epsilon^{2}$, where $\epsilon$ is the order of magnitude
of the initial perturbation.

Direct numerical simulations of the Euler equations close to the Kolmogorov
base flow show an excellent agreement with the above theoretical predictions.\\

\begin{figure}[H]

\begin{centering}
\includegraphics[width=0.44\textwidth]{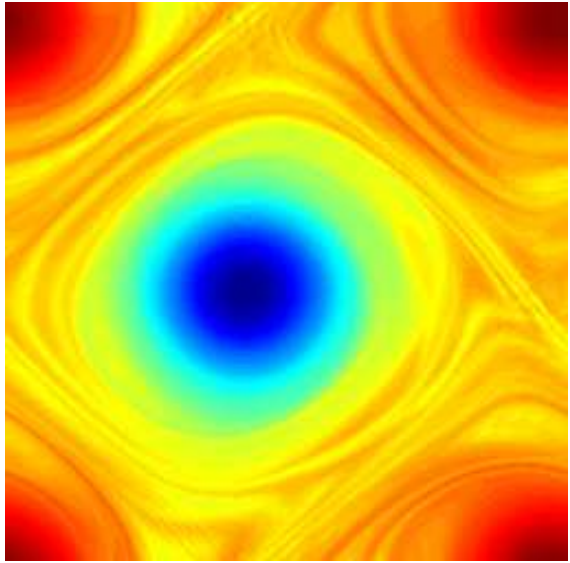}
\par\end{centering}

\caption{\label{fig:Vorticite-SNS}A snapshot of the vorticity field for the
2D Navier-Stokes equations with stochastic forces, in a statistically
stationary regime. The vorticity field is close to a steady state
of the Euler equation (here a dipole). The fluctuations close to this
state are the visible filaments on the figure. One clearly see that
such filaments are present in between the two vortices, but are absent
in the core of the vortices. This is due to the vorticity depletion
mechanism at the core of the vortices, the points where the angular
velocity of the vortices have local extrema.}

\end{figure}

The theoretical study performed in this paper, could be easily generalized
to the study of the asymptotic behavior and stability of jets in the
context of both barotropic flows in the beta-plane approximations,
or two dimensional axisymmetric vortices. Many recent works have considered
perturbations to two dimensional vortices \cite{Turner_Gilbert_2007,Bajer_Bassom_Gilbert_2001,Balmforth_Smith_Young_2001JFM...426...95B,Scecter_etal_2000_PhysicsFluids,Schecter_Montgommery_2003DyAtO..37...55S}.
As far as asymptotic behavior is concerned, following this paper approach,
in the case of vortices, we argue that a similar perturbation vorticity
depletion should occur at any stationary point of the angular velocity
of the vortex $U\left(r\right)/r$. For instance in the case of a
monotonic angular velocity, such a stationary point of the angular
velocity is located at the core of the vortex $r=0$, and the vorticity
depletion occurs at the center of the vortex. This phenomena has indeed
been observed by Bassom and Gilbert \cite{Bassom_Gilbert_JFM_1998}
(see their discussion, and the comment of their figure 2(b) and 4(a)).
They have stated that {}``We at present lack a simple physical explanation
of this process whereby vorticity is more highly suppressed than a
passive scalar, and do not know whether it has applicability beyond
the Gaussian vortex''. The type of arguments developed in the present
paper, based on the Laplace tools, suggests that such a vorticity
depletion is indeed a generic phenomenon, valid for any parallel flow
(resp. circular vortex), at the stationary points of the velocity
profile (resp. angular velocity profile). Mathematically this is due
to the regularization of the critical layer singularities at the edge
of the continuous spectrum.

This vorticity depletion mechanism also impacts turbulent flows where
the perturbations are locally governed by the linearized equations.
Fig \ref{fig:Vorticite-SNS} shows a snapshot of the vorticity field
in the dynamics of the stochastic 2D Navier-Stokes equations \cite{Bouchet_Simonnet_2008}.
One clearly observes a depletion of the vorticity fluctuations at
the core of the vortices. This effect in a stochastically forced equation
is related to the depletion mechanism in a deterministic equation,
described in this paper.\\

We have treated the linearized dynamics and the asymptotic stability
for the case of parallel flows, for the 2D Euler dynamics. The generalization
of such results for more complex cases, for instance flow with separatrix
and stationary points would also be extremely interesting. The problem
is then much more difficult from a theoretical point of view, but
could be addressed numerically. Also the time dependent situation,
by contrast to the case of perturbation of steady base flows, is of
a large interest. It has been shown numerically that interactions
with large scales dominate the small scale dynamics \cite{Dubrulle_Nazarenko_1997PhyD,Laval_Dubrulle_Nazarenko_JCompPhys_2004},
in the spirit of rapid distortion theory or quasilinear approaches.
This has deep impact on the statistics of the associated turbulence
\cite{Nazarenko_Laval_JFM_2000}.\\

Because both are transport equations by a non-divergent field, there
is a very well known analogy between the Vlasov equations and the
2D Euler equations. An even closer relation between the 2D-Euler and Vlasov equation can also be found: the 2D -Euler dynamics of a localized perturbation (vorticity defect) is actually described by a Vlasov equation at leading order \cite{Balmforth_Del-Castillo-Negrete_Young_1997_JFM}. Both the 2D-Euler and Vlasov equations have very similar behaviors, including
for instance relaxations without dissipation (Orr mechanism or Landau
damping) and the associated asymptotic stability. The theory of the
asymptotic stability of 2D Euler equations is thus deeply related
to the asymptotic stability of Vlasov equations. We note very recent
mathematical results on nonlinear Landau damping \cite{Mouhot_Villani:2009},
a subject related to the asymptotic stability of Vlasov equations.
A natural issue, is to know if such recent mathematical results \cite{Mouhot_Villani:2009},
could be generalized to the case of the 2D Euler equations, in relation
with the results obtained in the present work.

It is important to notice that a rigorous mathematical treatment of
the Orr mechanism for the Euler equations, in the spirit of the work
by Mouhot and Villani \cite{Mouhot_Villani:2009} for the Vlasov equation,
does not exist yet, even in the simplest case of base flow profile $U$ without
stationary point. This is a very interesting mathematical problem
and we hope that some new results may follow the recent proof in the
Vlasov case.

The Laplace tools we have used along this paper is suited for
analytical initial data only. An interesting question is to know if
similar results may hold for non analytic data. We note that for the
Vlasov equations, analyticity of initial data leads to exponential
decay of the perturbation ; but there are counterexamples by Glassey
and Schaeffer \cite{Glassey_Schaeffer_1994,Glassey_Schaeffer_1995} showing
that there is in general no exponential decay for the linearized Vlasov Poisson
equation without analyticity, or without confinement. This seems to
indicate that analyticity is essential for observing Landau damping
for the Vlasov equation. We refer to \cite{Mouhot_Villani:2009}
for a further discussion of this point. We guess that similarly, analyticity
is an essential ingredient for the Orr mechanism in the context of
the 2D Euler equations.

The current work has interesting consequences for the understanding
of the kinetic theory of the point vortex model in one hand and for the linearized 2D Euler and 2D Navier Stokes equations with stochastic
forces, when the large scale structures dominate the dynamics on the other hand. These
applications were one of our motivation for studying the asymptotic behavior of
the linearized Euler equation and asymptotic stability of the Euler
equations. These consequences will be developed in forthcoming works.

\section*{Acknowledgments}

We warmly thank J. Barré and E. Simonnet for useful discussions about
this work.

This work was supported through the ANR program STATFLOW (ANR-06-JCJC-0037-01) and through the ANR program STATOCEAN (ANR-09-SYSC-014).

\newpage{}

\section*{A) Oscillating integrals}

\subsection*{A-1) General classical results}

Let us consider the asymptotic behavior, for large $t$, of the integral
\begin{equation}
I(t)=\int_{a}^{b}dx\, g(x)\exp(if(x)t)\label{eq:Oscillating integral}\end{equation}

\begin{enumerate}

\item
First, we consider the case when $f$ has no singular point.
We suppose that $f$ and $g$ are twice differentiable,
that $f$ has no singular point, i.e. $f'(x)\neq0$ for any $x$,
and that either $g(a)\neq0$ or $g(b)\neq0$. Then \begin{equation}
I(t)\sim\frac{1}{it}\left[\frac{g(b)\exp\left(itf(b)\right)}{f'(b)}-\frac{g(a)\exp\left(itf(a)\right)}{f'(a)}\right]\label{eq:Oscillating-NonStationnaire}\end{equation}
Namely, $I(t)\sim t^{-1}$.
This can be easily proved by integrations by part.
If $g(a)=g(b)=0$, if $g$ and $f$ are sufficiently differentiable,
then the asymptotic behavior can be obtained by further integrations by parts.

\item
Next, we consider the case when $f$ has a single stationary point.
We suppose that $f$ is a real function with a single stationary
point $x_{0}$ ($f'(x_{0})=0$),
that $f$ and $g$ are smooth and that $f''\left(x_{0}\right)\neq0$.
The asymptotic behavior of the oscillating
integral (\ref{eq:Oscillating integral}) is then given by the classical
stationary phase results \cite{Erdelyi_1956,Bleistein_Handelsman_1975},
\begin{equation}
I(t)\underset{t\rightarrow\infty}{\sim}g(x_{0})\sqrt{\frac{2\pi}{\left|f''\left(x_{0}\right)\right|}}\exp\left(\frac{i\epsilon\pi}{4}\right)\frac{\exp\left(itf\left(x_{0}\right)\right)}{\sqrt{t}},\label{eq:Oscillating-Stationnaire}\end{equation}
 where $\epsilon$ is the sign of $f''\left(x_{0}\right)$.
Namely, $I(t)\sim t^{-1/2}$.
\end{enumerate}

\subsection*{A-2) Oscillating integrals and the velocity asymptotic expansion}

We apply the general methods of oscillating integrals to the particular
case of the computation of the velocity from an oscillating vorticity
field, like for instance the case given by equation (\ref{eq:velocity_ocillating_integral}).
We first evaluate the long time asymptotics of \begin{equation}
\mathbf{v}(y,t)=\int dy'\,\mathbf{G}_{k}(y,y')h(y')\exp(-ikU\left(y'\right)t),\label{eq:Evolution_velocity_2}\end{equation}
 where the function $h$ is assumed to be twice differentiable and
where $\mathbf{G}_{k}$ is the Green function for the computation
of the velocity $\mathbf{v}(y)\exp\left(ikx\right)$ from a vorticity
field $\omega\left(y\right)\exp\left(ikx\right)$. We treat explicitly
the case of a channel $-L<y<L$. The results are also valid for periodic
boundary conditions for $y$, just by dropping all the contributions
from the boundary in the expressions bellow.

Before going into more general discussions,
we note that for the case of a linear base flow $U\left(y\right)=sy$,
and when $h(y)$ is either constant, sinusoidal, or hyperbolic sinusoidal,
then an explicit expression for $\mathbf{v}_{k}(y,t)$ is
obtained (see \cite{Case_1960_Phys_Fluids}).

Clearly, equation (\ref{eq:Evolution_velocity_2}) is an oscillating integral.
In order to study its asymptotic expansion, we use the results of
section A-1 of this appendix. \\

First, we treat the case of a base flow without stationary point.
We use the fact that
$\mathbf{G}_{k}(y,.)$ is smooth everywhere, except for $y'=y$ (see
appendix B). Then we can use the results on the asymptotic behaviors of oscillating
integrals (section A-1) for both intervals $-L<y'<y$ and $y<y'<L$
independently. Moreover, we assume that $h\left(y\right)$ is at least twice
differentiable. We then obtain \begin{equation}
v_{x}(y,t)\underset{t\gg l/ks}{\sim}-\frac{1}{ikst}\left[G_{k,x}\left(y,L\right)h(L)\exp\left(-ikU\left(L\right)t\right)-G_{k,x}\left(y,-L\right)h(-L)\exp\left(ikU\left(-L\right)t\right)+h(y)\exp\left(-ikU\left(y\right)t\right)\right]\,;\label{eq:Response-velocity-x}\end{equation}
 the first two terms are contributions from the boundaries and the
third term is due to the discontinuity of $G_{k,x}\left(y,y'\right)$
for $y=y'$ (see (\ref{eq:deltaG1}) in appendix B). Here we define
$s$ as the minimum shear rate $s=\min(U'\left(y\right))$, and $l$
is a typical length scale that characterizes the variations of $h\left(y\right)$.
Similarly we obtain \begin{equation}
v_{y}(y,t)\underset{t\gg l/ks}{\sim}\frac{1}{k^{2}s^{2}t^{2}}\left[\frac{\partial G_{k,y}}{\partial y'}\left(y,L\right)h\left(L\right)\exp\left(-ikU\left(L\right)t\right)-\frac{\partial G_{k,y}}{\partial y'}\left(y,-L\right)h\left(-L\right)\exp\left(ikU\left(L\right)t\right)-ikh\left(y\right)\exp\left(-ikU\left(y\right)t\right)\right]\,;\label{eq:Response-velocity-y}\end{equation}
 We note that there is no contribution of order $1/t$ in this case, because $G_{k,y}(y,y')$
has a discontinuity of its first derivative only, for $y'=y$ (see
(\ref{eq:deltaG2}) in appendix B ; moreover  in the case of a bounded domain $G_{k}$ vanishes at the boundaries ($G_{k,y}(y,L)=0$)).\\

Next, we treat the case of a base flow with stationary points $y_{m}$
such that $U'\left(y_{m}\right)=0$. We assume that each stationary
point $y_{m}$ is not degenerated ($f''(y_{m})\neq0$). We perform
the computation for only one of these, denoted $y_{0}$,
without loss of generality.

Recall that the classical results on oscillating integrals (\ref{eq:Oscillating-Stationnaire})
assume the function $g$ (see Eq. \ref{eq:Oscillating integral})
to be smooth. This is not always the case for us. Indeed, the Green
function $G_{k}(y,y')$ is not smooth for $y=y'$. However, if $y\neq y_{0}$,
this discontinuity for $G_{k}$ or for its derivative can easily
be handled by dividing the integration interval into two subintervals,
as has been done in the previous paragraph. Then we conclude that
the leading order of the asymptotic behavior is still dominated by
the contribution of the stationary points. Thus the result (\ref{eq:Oscillating-Stationnaire})
is still valid.

Then, from (\ref{eq:Evolution_velocity_2}), using (\ref{eq:Oscillating-Stationnaire}),
we obtain, for $y\neq y_{0}$, \begin{equation}
\mathbf{v}(y,t)\underset{t\rightarrow\infty}{\sim}\mathbf{G}_{k}(y,y_{0})h(y_{0})\sqrt{\frac{2\pi}{\left|kU''\left(y_{0}\right)\right|}}\exp\left(\frac{i\epsilon_{k}\pi}{4}\right)\frac{\exp(-ikU\left(y_{0}\right)t)}{\sqrt{t}},\label{eq:vk-stationnaire}\end{equation}
 where $\epsilon_{k}$ is the sign of $-kU''\left(y_{0}\right)$.
We note that the asymptotic expansion has a discontinuity for $y=y_{0}$,
due to the discontinuity of the Green function.
Through a straightforward
generalization of the classical results (\ref{eq:Oscillating-Stationnaire})
to oscillating integrals with discontinuous functions $g$,
we can conclude that this discontinuity is regularized over a length scale
$l=\sqrt{1/(\left|kU''(y_{0})\right|t)}$, that decreases with time.

\subsection*{B) Green functions }

Let us establish the expression and some properties for the velocity
Green function $\mathbf{G}_{k}$. The results on the discontinuity
of the Green function, equations (\ref{eq:deltaG1}) and (\ref{eq:deltaG2})
bellow, are necessary for the discussion of appendix A.
We here consider the case of periodic boundary conditions ($y$ $2\pi-$periodic),
though the case of a channel geometry $-L\leq y\leq L$ can be treated similarly,
the resultant equations (\ref{eq:deltaG1}) and (\ref{eq:deltaG2}) remaining unchanged.

Let us denote $H_{k}$ the Green function for the stream function
in the $x$-Fourier space.
The $x$-Fourier transforms of $\omega=\Delta\psi$ gives
$\omega_{k}=d^{2}\psi_{k}/dy^{2}-k^{2}\psi_{k}$.
$H_{k}(y,y')$ is thus solution of \begin{equation}
\frac{\partial^{2}H_{k}}{\partial y^{2}}-k^{2}H_{k}=\delta\left(y-y'\right)\label{eq:Hk}\end{equation}
where $H_{k}(.,y')=0$ is a $2\pi-$periodic function for any $y$.

We note that $H_{k}\left(y,y'\right)$, considered as function of
$y$, is smooth everywhere except for $y=y'$. For $y=y'$, $H_{k}$
is continuous and has a jump unity for its first derivative:
\begin{equation}
\frac{\partial H_{k}}{\partial y}\left(y'^{+},y'\right)-\frac{\partial H_{k}}{\partial y}\left(y'^{-},y'\right)=1,\label{eq:derivatives_Hk}
\end{equation}
 where $F\left(y'^{+},y'\right)$ is the limit of $F(y,y')$ for
$y'$ going to $y$ with the condition $y>y'$.

Because of the translational invariance in a doubly periodic domain, the set of equations (\ref{eq:Hk}-\ref{eq:derivatives_Hk}) and the associate boundary conditions are invariant under translation. Then clearly, $H_{k}\left(y,y'\right)$ depends only on $||y-y'||$ where
\[
 ||y|| = \min_{{\rm integer}\ n}\left| 2\pi n-y \right|.
\]

Besides these general properties, an explicit expression to $H_{k}$
can be found from (\ref{eq:Hk}) and (\ref{eq:derivatives_Hk}):\[
H_k(y,y') = - \frac{\cosh (k||y-\pi||)}{2\sinh (k\pi)}
\]
Using the fact that $H_{k}\left(y,y'\right)$ depends only on $||y-y'||$ and (\ref{eq:derivatives_Hk}), it is easily verified that $H_{k}\left(y,y'\right)$,
considered as a function of $y'$, is differentiable and has a discontinuity
in its derivative for $y'=y$:\[
\frac{\partial H_{k}}{\partial y}\left(y,y^{+}\right)-\frac{\partial H_{k}}{\partial y}\left(y,y^{-}\right)=1\]

Using $\mathbf{v}=\nabla\wedge\left(\psi\mathbf{e}_{z}\right)$,
we have for the $x$-Fourier transforms : $\mathbf{v}_{k,x}=-d\psi_{k}/dy$
and $\mathbf{v}_{k,y}=ik\psi_{k}$. Thus \[
\mathbf{G}_{k}=\left(-\frac{\partial H_{k}}{\partial y},ikH_{k}\right)\]
 Then, using the properties of $H_{k}$, we note that $\mathbf{G}_{k}\left(.,y'\right)$
is smooth everywhere except for $y=y'$,
and that its derivative has a
jump for $y=y'$ :
\begin{equation}
\mathbf{G}_{k}\left(y,y^{+}\right)-\mathbf{G}_{k}\left(y,y^{-}\right)=\left(-1,0\right)\label{eq:deltaG1}\end{equation}
and
\begin{equation}
\frac{\partial\mathbf{G}_{k,y}}{\partial y}\left(y,y^{+}\right)-\frac{\partial\mathbf{G}_{k,y}}{\partial y}\left(y,y^{-}\right)=ik\label{eq:deltaG2}\end{equation}

\newpage{}

\bibliographystyle{plain} \bibliographystyle{plain} \bibliographystyle{plain}
\bibliography{FBouchet,Long_Range,Meca_Stat_Euler,Experimental_2D_Flows,Euler2D-Linearized,Turbulence_2D,Quasilinear,Euler_Stability,Phase_Stationnaire,num,Kinetic-Theories-Turbulence}

\end{document}